\documentclass[12pt]{article}
\usepackage{enumerate}
\usepackage{amsfonts}
\usepackage{amsmath}
\usepackage{amssymb}
\usepackage{amsthm}
\usepackage{latexsym}
\usepackage{color}

\def\D{\mathcal{D}}
\def\H{\mathcal{H}}

\def\P{\mathfrak{P}}
\def\SC{\mathcal{S}}
\def\S{\mathfrak{S}}

\def\C{\mathfrak{C}}
\def\T{\mathfrak{T}}

\def\N{\mathbb{N}_0}

\newcommand{\supp}{\mathrm{supp}}
\newcommand{\rank}{\mathrm{rank}}

\newcommand{\id}{\mathrm{Id}}
\newcommand{\Tr}{\mathrm{Tr}}

\newcommand{\shs}{\hspace{1pt}}
\newcounter{defin}  \newcounter{lemma}  \newcounter{theorem}
\newcounter{proposition} \newcounter{corol}  \newcounter{remark} \newcounter{example}
\newenvironment{lemma}{\par\refstepcounter{lemma}     \textbf{Lemma \thelemma.} }{\rm\par}

\newenvironment{corollary}{\par\refstepcounter{corol}     \textbf{Corollary \thecorol.} }{\rm\par}

\begin{document}

\title{Local lower bounds on characteristics of quantum and classical systems}

\author{M.E.~Shirokov \\
Steklov Mathematical Institute, Moscow, Russia}
\date{}
\maketitle
\begin{abstract} We consider methods for obtaining local lower bounds on characteristics of quantum (correspondingly, classical) systems, i.e. lower bounds
valid in the \break $\|\cdot\|_1\textup{-}$norm $\epsilon$-neighborhood of a given state (correspondingly, probability distribution). The main attention is paid to
infinite-dimensional systems.
\end{abstract}

\tableofcontents

\pagebreak
\section{Introduction}

There are many characteristics used in quantum (resp. classical)  information theory that are nonnegative
lower semicontinuous functions either on the set of all quantum states (resp. probability distributions)
or on some its proper subset. The list of basic examples includes:
\begin{itemize}
   \item the von Neumann entropy  (resp. the Shannon entropy);
   \item the quantum relative entropy  (resp. the Kullback-Leibler divergence);
   \item the quantum conditional entropy of q-c states (resp. equivocation);
   \item the quantum  (resp. classical)  conditional mutual information;
   \item several entanglement measures (the entanglement of formation, the relative entropy of entanglement, the squashed entanglement).
\end{itemize}

If a function $f$ defined on a subset $\S_0\subseteq\S(\H)$ is lower semicontinuous at a state $\rho\in \S_0$
then it is easy to see that the quantity
\begin{equation*}
L_f(\rho\shs|\shs\varepsilon)=\inf\left\{f(\sigma)\,|\,\sigma\in\S_0,\, \textstyle\frac{1}{2}\|\rho-\sigma\|_1\leq\varepsilon\right\}
\end{equation*}
tends to $f(\rho)\leq+\infty\,$ as $\,\varepsilon\to 0$. Naturally, the question arises about the estimation of this quantity. We will say that
$\widehat{L}_f(\rho\shs|\shs\varepsilon)$ is a \emph{faithful local lower bound} on a function $f$ at a state $\rho$ if
$$
\widehat{L}_f(\rho\shs|\shs\varepsilon)\leq L_f(\rho\shs|\shs\varepsilon)\quad \forall\varepsilon\in(0,1]\quad\textrm{and}\quad \lim_{\varepsilon\to0}\widehat{L}_f(\rho\shs|\shs\varepsilon)=f(\rho)\leq+\infty.
$$

In this article we consider general ways of getting faithful local lower bounds on basic characteristics of quantum systems
and discrete random variables.

\section{Preliminaries}

\subsection{Quantum systems}

Let $\mathcal{H}$ be a separable Hilbert space,
$\mathfrak{B}(\mathcal{H})$ the algebra of all bounded operators on $\mathcal{H}$ with the operator norm $\|\cdot\|$ and $\mathfrak{T}( \mathcal{H})$ the
Banach space of all trace-class
operators on $\mathcal{H}$  with the trace norm $\|\!\cdot\!\|_1$. Let
$\mathfrak{S}(\mathcal{H})$ be  the set of quantum states (positive operators
in $\mathfrak{T}(\mathcal{H})$ with unit trace) \cite{H-SCI,N&Ch,Wilde}.

Write $I_{\mathcal{H}}$ for the unit operator on a Hilbert space
$\mathcal{H}$ and $\id_{\mathcal{\H}}$ for the identity
transformation of the Banach space $\mathfrak{T}(\mathcal{H})$.

We will use the Mirsky inequality
\begin{equation}\label{Mirsky-ineq+}
  \sum_{i=1}^{+\infty}|\lambda^{\rho}_i-\lambda^{\sigma}_i|\leq \|\rho-\sigma\|_1
\end{equation}
valid for any positive operators $\rho$ and $\sigma$ in $\T(\H)$, where  $\{\lambda^{\rho}_i\}_{i=1}^{+\infty}$
and $\{\lambda^{\sigma}_i\}_{i=1}^{+\infty}$ are the sequences
of eigenvalues of $\rho$ and $\sigma$ arranged in the non-increasing order (taking the multiplicity into account) \cite{Mirsky,Mirsky-rr}.

The \emph{von Neumann entropy} of a quantum state
$\rho \in \mathfrak{S}(\H)$ is  defined by the formula
$S(\rho)=\operatorname{Tr}\eta(\rho)$, where  $\eta(x)=-x\ln x$ if $x>0$
and $\eta(0)=0$. It is a concave lower semicontinuous function on the set~$\mathfrak{S}(\H)$ taking values in~$[0,+\infty]$ \cite{H-SCI,L-2,W}.
The von Neumann entropy satisfies the inequality
\begin{equation}\label{w-k-ineq}
S(p\rho+(1-p)\sigma)\leq pS(\rho)+(1-p)S(\sigma)+h_2(p)
\end{equation}
valid for any states  $\rho$ and $\sigma$ in $\S(\H)$ and $p\in[0,1]$, where $\,h_2(p)=\eta(p)+\eta(1-p)\,$ is the binary entropy \cite{O&P,N&Ch,Wilde}.
Note that inequality (\ref{w-k-ineq}) holds with $h_2(p)$ replaced by the \emph{nondecreasing concave} continuous function $\tilde{h}_2(p)$ on $[0,1]$ defined as
\begin{equation}\label{h+}
\tilde{h}_2(p)=\left\{\begin{array}{l}
        h_2(p)\;\; \textrm{if}\;\;  p\in[0,\frac{1}{2}]\\
        \ln2\;\quad \textrm{if}\;\;  p\in(\frac{1}{2},1].
        \end{array}\right.
\end{equation}

We will use the  homogeneous extension of the von Neumann entropy to the positive cone $\T_+(\H)$ defined as
\begin{equation}\label{S-ext}
S(\rho)\doteq(\Tr\rho)S(\rho/\Tr\rho)=\Tr\eta(\rho)-\eta(\Tr\rho)
\end{equation}
for any nonzero operator $\rho$ in $\T_+(\H)$ and equal to $0$ at the zero operator \cite{L-2}.

By using concavity of the von Neumann entropy and inequality (\ref{w-k-ineq}) it is easy to show that
\begin{equation}\label{w-k-ineq+}
S(\rho)+S(\sigma)\leq S(\rho+\sigma)\leq S(\rho)+S(\sigma)+H(\{\Tr\rho,\Tr\sigma\})
\end{equation}
for any $\rho$ and $\sigma$ in $\T_+(\H)$, where  $H(\{\Tr\rho,\Tr\sigma\})=\eta(\Tr\rho)+\eta(\Tr\sigma)-\eta(\Tr(\rho+\sigma))$
is the homogeneous extension of the binary entropy to the positive cone in $\mathbb{R}^2$.

Let $H$ be a positive (semi-definite)  operator on a Hilbert space $\mathcal{H}$ (we will always assume that positive operators are self-adjoint). Write  $\mathcal{D}(H)$ for the domain of $H$. For any positive operator $\rho\in\T(\H)$ we will define the quantity $\Tr H\rho$ by the rule
\begin{equation}\label{H-fun}
\Tr H\rho=
\left\{\begin{array}{l}
        \sup_n \Tr P_n H\rho\;\; \textrm{if}\;\;  \supp\rho\subseteq {\rm cl}(\mathcal{D}(H))\\
        +\infty\;\;\textrm{otherwise}
        \end{array}\right.
\end{equation}
where $P_n$ is the spectral projector of $H$ corresponding to the interval $[0,n]$ and ${\rm cl}(\mathcal{D}(H))$ is the closure of $\mathcal{D}(H)$.\footnote{The support $\mathrm{supp}\rho$ of a state $\rho$ is the closed subspace spanned by the eigenvectors of $\rho$ corresponding to its positive eigenvalues.} If
$H$ is the Hamiltonian (energy observable) of a quantum system described by the space $\H$ then
$\Tr H\rho$ is the mean energy of a state $\rho$.

For any positive operator $H$ the set
$$
\C_{H,E}=\left\{\rho\in\S(\H)\,|\,\Tr H\rho\leq E\right\}
$$
is convex and closed (since the function $\rho\mapsto\Tr H\rho$ is affine and lower semicontinuous). It is nonempty if $E> E_0$, where $E_0$ is the infimum of the spectrum of $H$.

The von Neumann entropy is continuous on the set $\C_{H,E}$ for any $E> E_0$ if and only if the operator $H$ satisfies  the \emph{Gibbs condition}
\begin{equation}\label{H-cond}
  \Tr\, e^{-\beta H}<+\infty\quad\textrm{for all}\;\,\beta>0
\end{equation}
and the supremum of the entropy on this set is attained at the \emph{Gibbs state}
\begin{equation*}
\gamma_H(E)\doteq e^{-\beta(E) H}/\Tr e^{-\beta(E) H},
\end{equation*}
where the parameter $\beta(E)$ is determined by the equation $\Tr H e^{-\beta H}=E\Tr e^{-\beta H}$ \cite{W}. Condition (\ref{H-cond}) can be valid only if $H$ is an unbounded operator having  discrete spectrum of finite multiplicity. It means, in Dirac's notation, that
\begin{equation}\label{H-form}
H=\sum_{k=0}^{+\infty} E_k |\tau_k\rangle\langle\tau_k|,
\end{equation}
where
$\mathcal{T}\doteq\left\{\tau_k\right\}_{k=0}^{+\infty}$ is the orthonormal
system of eigenvectors of $H$ corresponding to the \emph{nondecreasing} unbounded sequence $\left\{E_k\right\}_{k=0}^{+\infty}$ of its eigenvalues
and \emph{it is assumed that the domain $\D(H)$ of $H$ lies within the closure $\H_\mathcal{T}$ of the linear span of $\mathcal{T}$}. In this case
\begin{equation*}
\Tr H \rho=\sum_i \lambda_i\|\sqrt{H}\varphi_i\|^2
\end{equation*}
for any operator $\rho$ in $\T_+(\H)$ with the spectral decomposition $\rho=\sum_i \lambda_i|\varphi_i\rangle\langle\varphi_i|$ provided that
all the vectors $\varphi_i$ lie in $\D(\sqrt{H})=\{ \varphi\in\H_\mathcal{T}\,| \sum_{k=0}^{+\infty} E_k |\langle\tau_k|\varphi\rangle|^2<+\infty\}$. If at least one eigenvector of $\rho$ corresponding to a nonzero eigenvalue does not belong to the set $\D(\sqrt{H})$
then $\Tr H \rho=+\infty$.


We will use the function
\begin{equation}\label{F-def}
F_{H}(E)\doteq\sup_{\rho\in\C_{H,E}}S(\rho)=S(\gamma_H(E))=\beta(E)E+\ln \Tr e^{-\beta(E)H}.
\end{equation}
This is a strictly increasing concave function on $[E_0,+\infty)$ \cite{EC,W-CB}. It is easy to see that $F_{H}(E_0)=\ln m(E_0)$, where $m(E_0)$ is the multiplicity of $E_0$. By Proposition 1 in \cite{EC} the Gibbs condition (\ref{H-cond}) is equivalent to the following asymptotic property
\begin{equation}\label{H-cond-a}
  F_{H}(E)=o\shs(E)\quad\textrm{as}\quad E\rightarrow+\infty.
\end{equation}

For example, if $\,\hat{N}\doteq a^\dag a\,$ is the number operator of a quantum oscillator then $F_{\hat{N}}(E)=g(E)$, where
\begin{equation}\label{g-def}
 g(x)=(x+1)\ln(x+1)-x\ln x,\;\, x>0,\quad g(0)=0.
\end{equation}

We will often assume that
\begin{equation}\label{star}
  E_0\doteq\inf\limits_{\|\varphi\|=1}\langle\varphi|H|\varphi\rangle=0.
\end{equation}
In this case the function $F_H$ satisfies the conditions of the following lemma (cf.\cite[Corollary 12]{W-CB}).\smallskip

\begin{lemma}\label{W-L} \emph{If $f$ is a nonnegative concave function on $\mathbb{R}_+$ then}
\begin{equation*}
  xf(z/x)\leq yf(z/y)\quad \textit{ for all }\;y>x>0\;\textit{ and }\;z\geq0.
\end{equation*}
\end{lemma}

\subsection{Discrete random variables}

Any vector $X_1$,...,$X_n$  of random variables, each of which takes a countable set of values, is described by
$n$-variate probability distribution $\bar{p}=\{p_{i_1...i_n}\}_{(i_1,..,i_n)\in\N^n}$, where $\N$ is the set of non-negative integers.\footnote{We will assume that the values of a discrete random variable is indexed by the set $\N=\mathbb{N}\cup\{0\}$. This simplifies the transition from the case of quantum systems to classical ones.}
We will denote by $\P_n$ the set of all such  probability distributions equipped with the \emph{total variation
distance} $\mathrm{TV}$,  which is defined for any  $\bar{p}=\{p_{i_1...i_n}\}$ and $\bar{q}=\{q_{i_1...i_n}\}$ in $\P_n$ as follows
\begin{equation}\label{TVD}
\mathrm{TV}(\bar{p},\bar{q})\doteq\frac{1}{2}\displaystyle\sum_{i_1,..,i_n}|p_{i_1...i_n}-q_{i_1...i_n}|.
\end{equation}

For any $\bar{p}\in\P_n$ we will write $\bar{p}_k$ and $[\bar{p}_k]_i$ for the marginal distribution of $X_k$
and its $i$-th entry, i.e.
$$
\bar{p}_k=\{[\bar{p}_k]_i\}_i,\quad [\bar{p}_k]_i=\sum_{i_1,..,i_{k-1},i_{k+1},..,i_n}p_{i_1...i_{k-1}i\,i_{k+1}...i_n}.
$$
The number of nonzero entries of the distribution  $\bar{p}_k$ will be denoted by $|\bar{p}_k|$.

The \emph{Shannon entropy} of a probability distribution $\bar{p}=\{p_{i_1...i_n}\}\in\P_n$ is defined as
$$
H(\bar{p})=\sum_{i_1,..,i_n}\eta(p_{i_1...i_n}),
$$
where  $\eta(x)=-x\ln x$ if $x>0$ and $\eta(0)=0$. It is a concave lower semicontinuous function on the set~$\P_n$ taking values in~$[0,+\infty]$ \cite{C&T,H-SCI}.
The Shannon  entropy satisfies the inequality
\begin{equation}\label{w-k-ineq-c}
H(\lambda\bar{p}+(1-\lambda)\bar{q})\leq \lambda H(\bar{p})+(1-\lambda)H(\bar{q})+h_2(\lambda)
\end{equation}
valid for any distributions  $\bar{p}$ and $\bar{q}$ in $\P_n$ and $\lambda\in(0,1)$, where $\,h_2(\lambda)=\eta(\lambda)+\eta(1-\lambda)\,$ is the binary entropy \cite{O&P,N&Ch,Wilde}.


Let $\SC=\{E_i\}_{i=0}^{+\infty}$ be a nondecreasing sequence of nonnegative numbers such that
\begin{equation}\label{Z-cond}
\sum_{i=0}^{+\infty}e^{-\beta E_i}<+\infty\quad \forall\beta>0.
\end{equation}
Consider the function
\begin{equation}\label{F-Z}
\!F_\SC(E)=\sup\left\{H(\{p_i\})\,\left|\,\{p_i\}\in \P_1,\; \sum_{i=0}^{+\infty} E_ip_i\leq E\right.\right\}=\beta(E)E+\ln\sum_{i=0}^{+\infty} e^{-\beta(E)E_i},
\end{equation}
where $\beta(E)$ is defined by the equation $\sum_{i=0}^{+\infty} E_ie^{-\beta E_i}=E\sum_{i=0}^{+\infty} e^{-\beta E_i}$ \cite{W},\cite[Proposition 1]{EC}.

It is easy to see that the function $F_\SC(E)$ coincides with the function $F_H(E)$ defined in (\ref{F-def})
provided that $H$ is a positive operator with the spectrum $\SC$. If $\SC=\{0,1,2,...\}$ then $F_\SC(E)=g(E)$ -- the function defined in (\ref{g-def}). So,
Proposition 1 in \cite{EC} shows that condition (\ref{Z-cond}) is equivalent to the asymptotic property
\begin{equation}\label{Z-cond+}
F_\SC(E)=o(E)\quad \textrm{as}\quad E\to+\infty.
\end{equation}

\section{Basic lemma}

In this section we describe one general result concerning properties of  a real-valued function $f$ on a convex subset $\S_0$ of $\S(\H)$ satisfying the inequalities
\begin{equation}\label{LAA-1}
  f(p\rho+(1-p)\sigma)\geq pf(\rho)+(1-p)f(\sigma)-a_f(p)
\end{equation}
and
\begin{equation}\label{LAA-2}
  f(p\rho+(1-p)\sigma)\leq pf(\rho)+(1-p)f(\sigma)+b_f(p)
\end{equation}
for all states $\rho$ and $\sigma$ in $\S_0$ and any $p\in[0,1]$, where $a_f$ and $b_f$ are continuous  functions on $[0,1]$ such that $a_f(0)=b_f(0)=0$.
We will call functions
satisfying both inequalities (\ref{LAA-1}) and (\ref{LAA-2}) \emph{locally almost affine} (briefly, \emph{LAA functions}). We will assume that $a_f(p)=a_f(1-p)$ and $b_f(p)=b_f(1-p)$ for all $p\in[0,1]$. For technical simplicity we will also assume that
\begin{equation}\label{a-b-assump}
 \textrm{ the functions}\;\; a_f\;\; \textrm{and}\;\; b_f\;\; \textrm{are concave and non-decreasing on}\;\; \textstyle[0,\frac{1}{2}].
\end{equation}

For any state $\rho$ with the spectral representation $\,\rho=\sum_i\lambda_i^{\rho}|\varphi_i\rangle\langle\varphi_i|\,$
and arbitrary $\varepsilon>0$ introduce the trace class positive operators
\begin{equation}\label{2-op}
 \rho\wedge\varepsilon I_{\H}\doteq \sum_i \min\{\lambda_i^{\rho},\varepsilon\}|\varphi_i\rangle\langle\varphi_i|,\quad
[\rho-\varepsilon I_{\H}]_+\doteq \sum_i \max\{\lambda_i^{\rho}-\varepsilon,0\}|\varphi_i\rangle\langle\varphi_i|.
\end{equation}
It is clear that $\rho=\rho\wedge\varepsilon I_{\H}+[\rho-\varepsilon I_{\H}]_+$ and that $\,\rho\wedge\varepsilon I_{\H}\to0\,$ in the trace norm as $\,\varepsilon\to0$.

For any function $f$ on a convex subset $\S_0$ of $\,\S(\H)$ we will use its  homogeneous extension $\tilde{f}$ to the cone $\widetilde{\S}_0$ generated by $\S_0$. It is defined as
\begin{equation}\label{G-ext}
\tilde{f}(\rho)\doteq(\Tr\rho)f(\rho/\Tr\rho),\quad \tilde{f}(0)=0.
\end{equation}

The following lemma allows us to obtain local lower bounds for  nonnegative LAA functions  valid for commuting states.\smallskip

\begin{lemma}\label{b-lemma}  \emph{Let $f$ be a function on a convex subset $\S_0$ of $\,\S(\H)$ taking values in $[0,+\infty]$ and satisfying inequalities (\ref{LAA-1}) and (\ref{LAA-2}) on $\S_0$ with possible values $+\infty$ in both sides. Let $\rho$ be a state in $\,\S_0$ with the properties
\begin{equation}\label{S-prop+}
\!(\mathrm{a})\quad  h(\rho)\in\widetilde{\S}_0\;\textrm{ for any  function }\;h:\mathbb{R}_+\to\mathbb{R}_+\textrm{ such that  }\, h(x)\leq x\;\,\forall x\in\mathbb{R}_+,
\end{equation}
\begin{equation}\label{S-prop++}
\!(\mathrm{b})\quad  [\rho-\sigma]_\pm\in\widetilde{\S}_0\;\textrm{ for any state }\sigma\textrm{ in }\,\S_0\,\textrm{ such that }\,[\rho,\sigma]=0\,\textrm{ and }\, \sigma\neq\rho,\quad
\end{equation}
where $[\rho-\sigma]_\pm$ are the positive and negative parts of the Hermitian operator  $\rho-\sigma$, $[\rho,\sigma]\doteq\rho\sigma-\sigma\rho$.}\medskip\pagebreak

\noindent A) \emph{If $\,f(\rho)<+\infty\,$ then
\begin{equation}\label{g-ob-c+0}
f(\sigma)\geq f(\rho)-\tilde{f}(\rho\wedge\varepsilon I_{\H})-D_f(\varepsilon)-\tilde{a}_f(\varepsilon)
\end{equation}
for any state $\sigma$ in $\,\S_0$ such that $\,[\rho,\sigma]=0\,$ and $\,\frac{1}{2}\|\rho-\sigma\|_1\leq\varepsilon\leq1$,
where $\rho\wedge\varepsilon I_{\H}$ is the operator defined in (\ref{2-op}), $\tilde{f}$ is the homogeneous extension of $f$ defined in (\ref{G-ext}),}
\begin{equation*}
 D_f(\varepsilon)=\displaystyle(1+\varepsilon)(a_f+b_f)\!\left(\frac{\varepsilon}{1+\varepsilon}\right)\quad \textit{and}\quad\tilde{a}_f(\varepsilon)=\left\{\begin{array}{l}
        a_f(\varepsilon)\;\; \shs\textrm{if}\;\;  \varepsilon\in\shs[0,\frac{1}{2}]\\
        a_f(\frac{1}{2})\;\; \textrm{if}\;\;  \varepsilon\in(\frac{1}{2},1].
        \end{array}\right.
\end{equation*}

\noindent B) \emph{If either $\,f(\rho)<+\infty\,$ or
\begin{equation}\label{f-a-c}
\!\tilde{f}([\rho-\varepsilon I_{\H}]_+)<+\infty\quad \forall\varepsilon\in(0,1],
\end{equation}
where $[\rho-\varepsilon I_{\H}]_+$ is the operator defined in (\ref{2-op}), then
\begin{equation}\label{g-ob-c+}
f(\sigma)\geq \tilde{f}([\rho-\varepsilon I_{\H}]_+)-D_f(\varepsilon)-\tilde{a}_f(\varepsilon)-a_f(1-r_{\varepsilon}),\quad r_{\varepsilon}=\Tr[\rho-\varepsilon I_{\H}]_+,
\end{equation}
for any state $\sigma$ in $\,\S_0$ such that $\,[\rho,\sigma]=0\,$ and $\,\frac{1}{2}\|\rho-\sigma\|_1\leq\varepsilon\leq1$.}
\end{lemma}\smallskip

\textbf{Remark 1.}  All the terms in the r.h.s. of (\ref{g-ob-c+0}) and (\ref{g-ob-c+}) are well defined and finite. Indeed,
the operators $\rho\wedge\varepsilon I_{\H}$ and $[\rho-\varepsilon I_{\H}]_+$ belong to the cone $\widetilde{\S}_0$ due to condition (\ref{S-prop+}).
If $\,f(\rho)<+\infty\,$ then inequality (\ref{LAA-1}) implies that $\tilde{f}(\rho\wedge\varepsilon I_{\H})<+\infty$ and $\tilde{f}([\rho-\varepsilon I_{\H}]_+)<+\infty$, since  $\rho\wedge\varepsilon I_{\H}+[\rho-\varepsilon I_{\H}]_+=\rho$.
\smallskip

\emph{Proof.} Everywhere further we will assume that $f(\sigma)<+\infty$, since otherwise (\ref{g-ob-c+0}) and (\ref{g-ob-c+}) hold trivially.
Assume that $\rho$ and $\sigma$ are commuting states in $\,\S_0$ such that $\,\frac{1}{2}\|\rho-\sigma\|_1=\epsilon\leq\varepsilon\leq1$.  Let $\{\varphi_k\}_{k=0}^{+\infty}$ be an orthonormal
basis such that
\begin{equation*}
\rho=\sum_{k=0}^{+\infty} \lambda^{\rho}_k |\varphi_k\rangle\langle \varphi_k|\quad \textrm{and}\quad \sigma=\sum_{k=0}^{+\infty} \lambda^{\sigma}_k |\varphi_k\rangle\langle \varphi_k|,
\end{equation*}
where $\{\lambda^{\rho}_k\}$ is the sequence of eigenvalues of $\rho$ \emph{arranged in the non-increasing order} and
$\{\lambda^{\sigma}_k\}$ is the corresponding sequence of eigenvalues of $\sigma$ (not nondecreasing, in general).

\smallskip

A) Following the Alicki-Fannes-Winter technique (cf.\cite{A&F,W-CB}) introduce the states
\begin{equation*}
\tau_+=\epsilon^{-1}\sum_{k=0}^{+\infty} [\lambda^{\rho}_k-\lambda^{\sigma}_k]_+ |\varphi_k\rangle\langle \varphi_k|\quad \textrm{and} \quad\tau_-=\epsilon^{-1}\sum_{k=0}^{+\infty} [\lambda^{\rho}_k-\lambda^{\sigma}_k]_- |\varphi_k\rangle\langle \varphi_k|,
\end{equation*}
where $\,[\lambda^{\rho}_k-\lambda^{\sigma}_k]_+=\max\{\lambda^{\rho}_k-\lambda^{\sigma}_k,0\}$ and $\,[\lambda^{\rho}_k-\lambda^{\sigma}_k]_-=\max\{\lambda^{\sigma}_k-\lambda^{\rho}_k,0\}$.
The states $\tau_+$ and $\tau_-$ belong to the set $\S_0$ due to the condition (\ref{S-prop++}). Then we have
\begin{equation*}
\frac{1}{1+\epsilon}\,\rho+\frac{\epsilon}{1+\epsilon}\,\tau_-=\omega_*=
\frac{1}{1+\epsilon}\,\sigma+\frac{\epsilon}{1+\epsilon}\,\tau_+,
\end{equation*}
where $\omega_*$ is a state in $\S_0$ (due to the convexity of $\S_0$). The obvious operator inequalities $\,\epsilon\tau_+\leq\rho\,$ and $\,\epsilon\tau_-\leq\sigma$ along with the assumed finiteness of $f(\rho)$ and $f(\sigma)$ imply, due to inequalities (\ref{LAA-1}) and (\ref{LAA-2}),
the finiteness of $f(\tau_+)$, $f(\tau_-)$ and $f(\omega_*)$.

By applying inequalities (\ref{LAA-1}) and (\ref{LAA-2}) to the above decompositions of $\omega_*$  we  obtain
$$
(1-p)(f(\rho)-f(\sigma))\leq p
(f(\tau_+)-f(\tau_-))+a_f(p)+b_f(p)
$$
where $p=\epsilon/(1+\epsilon)$. Hence
\begin{equation}\label{imp-ineq}
f(\rho)-f(\sigma)\leq \epsilon(f(\tau_+)-f(\tau_-))+D_f(\epsilon)\leq \epsilon f(\tau_+)+D_f(\varepsilon),
\end{equation}
where the last inequality follows from the nonnegativity of $f$ and the fact that the function $D_f$ is nondecreasing on $[0,1]$ due to the assumption (\ref{a-b-assump}).

Since $\epsilon\tau_+\leq\rho$ and $\tau_+\leq I_{\H}$, we have
\begin{equation}\label{cr-est}
 \epsilon\tau_+\leq \rho\wedge\epsilon I_{\H}\leq \rho\wedge\varepsilon I_{\H}.
\end{equation}
So,
by using inequality (\ref{LAA-1}) and taking the nonnegativity of $f$ into account we obtain
\begin{equation}\label{imp-ineq+}
\epsilon f(\tau_+)\leq \bar{r}_{\varepsilon}f(\bar{r}^{-1}_{\varepsilon}\rho\wedge\varepsilon I_{\H})+\bar{r}_{\varepsilon}a_f(\epsilon/\bar{r}_{\varepsilon})\leq \bar{r}_{\varepsilon}f(\bar{r}^{-1}_{\varepsilon}\rho\wedge\varepsilon I_{\H})+\tilde{a}_f(\varepsilon),
\end{equation}
where $\bar{r}_{\varepsilon}=1-r_{\varepsilon}=\Tr(\rho\wedge\varepsilon I_{\H})$ and the last inequality follows from Lemma \ref{W-L} and the assumption (\ref{a-b-assump}).
This inequality and (\ref{imp-ineq}) imply (\ref{g-ob-c+0}).\smallskip

B) Assume that  $f(\rho)<+\infty$. Since $\rho=\rho\wedge\varepsilon I_{\H}+[\rho-\varepsilon I_{\H}]_+$,
it follows from inequality (\ref{LAA-1}) that
$$
\bar{r}_{\varepsilon}f(\bar{r}^{-1}_{\varepsilon}\rho\wedge\varepsilon I_{\H})\leq f(\rho)-r_{\varepsilon}f(r^{-1}_{\varepsilon}[\rho-\varepsilon I_{\H}]_+)+a_f(1-r_{\varepsilon}).
$$
This inequality and (\ref{g-ob-c+0}) imply (\ref{g-ob-c+}).

Let $f(\rho)=+\infty$. Assume that (\ref{g-ob-c+}) does not hold for some $\varepsilon_0\in(0,1]$ and a state $\sigma_0$ diagonizable in the basis $\{\varphi_k\}$ such that $\,\frac{1}{2}\|\rho-\sigma_0\|_1=\epsilon_0\leq\varepsilon_0$.
Consider the sequence of states
\begin{equation*}
\rho_n=[\rho-\gamma_nI_{\H}]_+/\Tr[\rho-\gamma_nI_{\H}]_+,\quad \gamma_n=1/n,
\end{equation*}
well defined for all $n$ large enough. It is clear that $\,\rho_n\to\rho\,$ and, hence, $\,\epsilon_n \doteq\frac{1}{2}\|\rho_n-\sigma_0\|_1$ tends to $\epsilon_0$ as $n\to+\infty$.

Condition (\ref{f-a-c}) implies that $f(\rho_n)<+\infty$ for any $n$
and  hence (\ref{g-ob-c+}) holds with $\rho=\rho_n$ and $\,\varepsilon=\varepsilon_n\doteq\min\{\epsilon_n+\varepsilon_0-\epsilon_0,1\}\,$
by the above part of the proof. Since $\rho_n$ and $\varepsilon_n$ tend, respectively,  to $\rho$ and $\varepsilon_0$ as $n\to+\infty$, it is  easy to see that
the sequence of operators
$$
[\rho_n-\varepsilon_n I_{\H}]_+=c_n^{-1}[\rho-(c_n\varepsilon_n+\gamma_n)I_{\H}]_+,\quad c_n=\Tr[\rho-\gamma_nI_{\H}]_+,
$$
tends to the operator $\,[\rho-\varepsilon_0 I_{\H}]_+\,$ in the trace norm. Thus, since the functions $a_f$ and $b_f$
are continuous on $[0,1]$, to obtain a contradiction to the above assumption it suffices to show that
\begin{equation}\label{imp-l-r}
\lim_{n\to+\infty} f(r_{\varepsilon_n}^{-1}[\rho_n-\varepsilon_n I_{\H}]_+)=f(r_{\varepsilon_0}^{-1}[\rho-\varepsilon_0 I_{\H}]_+),
\end{equation}
where $r_{\varepsilon_n}=\Tr[\rho_n-\varepsilon_n I_{\H}]_+$ and $r_{\varepsilon_0}=\Tr[\rho-\varepsilon_0 I_{\H}]_+$. To prove this relation note first that condition (\ref{S-prop+}) implies that the set $\S_0$ contains the set $\S_{\rho}$ of states proportional to the spectral projectors of $\rho$ corresponding to its positive eigenvalues. Thus,  condition (\ref{f-a-c}) and  inequality (\ref{LAA-1}) allow us to show that the function $f$ takes finite values on the set $\S_{\rho}$. Let $n_0$ be a natural number such that $c_n\varepsilon_n\geq \varepsilon_0/2$ for all $n\geq n_0$, where $c_n=\Tr[\rho-\gamma_nI_{\H}]_+$. Then it is easy to see that
all the states proportional to the operators $\,[\rho_n-\varepsilon_n I_{\H}]_+=c_n^{-1}[\rho-(c_n\varepsilon_n+\gamma_n)I_{\H}]_+\,$, $n\geq n_0$, and $[\rho-\varepsilon_0 I_{\H}]_+$
belong to the convex hull of the \emph{finite} subset of $\S_{\rho}$ corresponding to the eigenvalues of $\rho$ which are not less than $\varepsilon_0/2$. By Corollary 3 in  \cite[Section 3.1.2]{QC}
the LAA-function $f$ is continuous on this convex hull. This implies (\ref{imp-l-r}). $\Box$\smallskip

\textbf{Remark 2.} Since the proof of Lemma \ref{b-lemma} uses the very coarse estimate (\ref{cr-est}), the
lower bounds given by this lemma can not be too accurate. The main advantage
of these lower bounds is their universality. Note also that the lower bound given by Lemma \ref{b-lemma}B
is easily computable as the operator  $[\rho-\varepsilon I_{\H}]_+$ has finite rank for any $\rho$ and $\varepsilon>0$. $\Box$\smallskip

Applications of Lemma \ref{b-lemma} to different characteristics of quantum systems are considered in Section 4.
The version of Lemma \ref{b-lemma} adapted to characteristics of discrete random variables is presented in Section 5.1 (Lemma \ref{b-lemma+}).

\section{Characteristics of quantum systems}

\subsection{The von Neumann entropy}


Since the von Neumann entropy is a continuous function  on the compact set of states of a
finite-dimensional quantum system, it attains its maximal and minimal values on the $\varepsilon$-neighbourhood $V_{\varepsilon}(\rho)\doteq\left\{\sigma\in\S(\H)\,|\,\textstyle\frac{1}{2}\|\rho-\sigma\|_1\leq\varepsilon\right\}\,$
of any state $\rho$. The states in $\,V_{\varepsilon}(\rho)$ at which these maximal and minimal values are attained
are described by Hanson  and Datta in \cite{H&D}.

In the infinite-dimensional case the von Neumann entropy is not bounded from above on  the $\varepsilon$-neighbourhood $V_{\varepsilon}(\rho)$
of any state $\rho$ for all $\varepsilon>0$, since it takes the value $+\infty$ on the dense subset of $\S(\H)$. But for any state $\rho$ in $\S(\H)$ we can consider
the quantity
\begin{equation*}
L_S(\rho\shs|\shs\varepsilon)=\inf\left\{S(\sigma)\,|\,\sigma\in\S(\H),\, \textstyle\frac{1}{2}\|\rho-\sigma\|_1\leq\varepsilon\right\},
\end{equation*}
where the infimum is not attained in general, since the $\varepsilon$-neighbourhood
of any state $\rho$ is not compact.
The lower semicontinuity of the von Neumann entropy implies that
$$
\lim_{\varepsilon\to0^+}L_S(\rho\shs|\shs\varepsilon)=S(\rho)\leq +\infty.
$$

The results of Section 4.1 in \cite{LCB} imply several
computable lower bounds on $L_S(\rho\shs|\shs\varepsilon)$ expressed via the spectrum of $\rho$.

Proposition 2 in \cite{LCB}  (proved by using  Audenaert's continuity bound \cite{Aud}) shows that
\begin{equation}\label{B-lb-1}
L_S(\rho\shs|\shs\varepsilon)\geq S(\rho)-\varepsilon\ln(\rank\rho-1)-h_2(\varepsilon)
\end{equation}
for any finite rank state $\rho$ and any $\varepsilon\leq 1-1/\rank\rho$. By Lemma 2.2 in \cite{H&D} the equality holds in (\ref{B-lb-1}) if and only if
the state  $\rho$ has the spectrum $(1-\varepsilon, \varepsilon/d,..., \varepsilon/d)$, where $d=\rank\rho-1$.

Proposition 1 in \cite{LCB}  shows that
\begin{equation}\label{B-lb-2}
 L_S(\rho\shs|\shs\varepsilon)\geq  S(\rho)-\varepsilon F_H(E_{H,\shs\varepsilon}(\rho)/\varepsilon)-g(\varepsilon)\geq S(\rho)-\varepsilon F_H(E/\varepsilon)-g(\varepsilon)
\end{equation}
for any state $\rho$ in $\S(\H)$ such that $\Tr H\rho=E<+\infty$, where $H$ is any positive operator on $\H$ satisfying conditions (\ref{H-cond}) and (\ref{star}), $F_H$ and $g$
are the functions defined in (\ref{F-def}) and (\ref{g-def}) correspondingly and\footnote{$\rho\wedge\varepsilon I_{\H}$  and $[\rho-\varepsilon I_{\H}]_+$ are the operators defined in (\ref{2-op}).}
$$
E_{H,\shs\varepsilon}(\rho)\doteq \Tr H(\rho\wedge\varepsilon I_{\H})=E-\Tr H[\rho-\varepsilon I_{\H}]_+=o(1)\quad\textrm{as }\;\;\varepsilon\to0.
$$
The lower bound (\ref{B-lb-2}) is faithful, since  condition (\ref{H-cond}) implies the convergence of the r.h.s. of (\ref{B-lb-2}) to $S(\rho)$ as $\varepsilon\to+\infty$ (due to the equivalence of (\ref{H-cond}) and (\ref{H-cond-a})). The last claim of Proposition 1 in \cite{LCB} implies that the lower bound (\ref{B-lb-2})
is quite accurate: for some states $\rho$ the gap between the l.h.s. and the r.h.s. of (\ref{B-lb-2}) is less than $g(\varepsilon)$.

The lower bound (\ref{B-lb-2}) is applicable
to any state  $\rho$ in $\S(\H)$ with finite $S(\rho)$, since for any such state there is a positive operator $H$ on $\H$ satisfying conditions (\ref{H-cond}) and (\ref{star}) such that $\Tr H\rho<+\infty$ \cite[Proposition 4]{EC}. The operator $H$ can be treated as a "free parameter", with which one can optimize lower bound (\ref{B-lb-2}).
The second claim of Proposition 1 in \cite{LCB} and the arguments from its  proof show that with this optimization, we can restrict attention to positive operators $H$
having the form (\ref{H-form}), in which $\{\tau_k\}$ is the basis of eigenvectors of the state $\rho$ corresponding to the nondecreasing
sequence $\{\lambda^{\rho}_k\}_{k=0}^{+\infty}$  of its eigenvalues. By taking such an operator $H$ we can rewrite (\ref{B-lb-2}) as follows
\begin{equation}\label{B-lb-2+}
 L_S(\rho\shs|\shs\varepsilon)\geq  S(\rho)-\varepsilon F_\SC(E_{\SC,\shs\varepsilon}(\rho))/\varepsilon)-g(\varepsilon)
\end{equation}
where $F_\SC(E)$ is the function defined in (\ref{F-Z}) via a non-decreasing sequences $\,\SC=\{E_k\}_{k=0}^{+\infty}$ of nonnegative numbers with $\,E_0=0\,$ satisfying condition (\ref{Z-cond}) such that $\,\sum_{k=0}^{+\infty}E_k\lambda^{\rho}_k=E<+\infty\,$ and
\begin{equation}\label{E-H+}
 E_{\SC,\shs\varepsilon}(\rho)\doteq \sum_{k=0}^{+\infty}E_k\min\{\lambda^{\rho}_k,\varepsilon\}=E-\sum_{k\in I_\varepsilon} E_k(\lambda^{\rho}_k-\varepsilon)=o(1)\quad\textrm{as }\;\;\varepsilon\to0,
\end{equation}
where $I_\varepsilon$ is the (finite) set of all $k$ such that $\lambda^{\rho}_k>\varepsilon$.\smallskip

\textbf{Example 1.} Let $\rho_N$ be the Gibbs state of one-mode quantum oscillator with the mean number of quanta $N$ that has the spectral representation
$$
\rho_N=(1-q)\sum_{k=0}^{+\infty}q^k|\tau_k\rangle\langle\tau_k|,\quad q=\frac{N}{N+1},
$$
where $\{\tau_k\}$ is the Fock basis \cite{H-SCI}. Let $\SC=(0,1,2,...,n,..)$. Then
\begin{equation*}
 E_{\SC,\shs\varepsilon}(\rho_N)=(1-q)\sum_{k=0}^{+\infty}kq^k=\frac{q}{(1-q)}=N\quad\textrm{if}\;\;\varepsilon(N+1)\geq 1
\end{equation*}
and
\begin{equation*}
 E_{\SC,\shs\varepsilon}(\rho_N)=\varepsilon\sum_{k=0}^{n_{\varepsilon}-1}k+\sum_{k=n_{\varepsilon}}^{+\infty}kq^k=\frac{\varepsilon n_{\varepsilon}(n_{\varepsilon}-1)}{2}+\frac{q^{n_{\varepsilon}}n_{\varepsilon}}{1-q} +\frac{q^{n_{\varepsilon}+1}}{(1-q)^2} \quad\textrm{if}\;\;\varepsilon(N+1)< 1,
\end{equation*}
where $\,n_{\varepsilon}=[\log_q(\varepsilon(N+1))]+1\,$ is the minimal natural number $n$ such that $(1-q)q^n<\varepsilon$. Thus, in the case $\varepsilon(N+1)< 1$ we have $E_{\SC,\shs\varepsilon}(\rho_N)\leq\varepsilon R_N(\varepsilon)=o(1)$ as $\varepsilon\to0$, where
$$
R_N(\varepsilon)\doteq \frac{A^2_N(\varepsilon)}{2}+(N+1)^2(A_N(\varepsilon)+N),\quad A_N(\varepsilon)=\log_\frac{N}{N+1}(\varepsilon(N+1))+1.
$$
Since $F_\SC(E)=S(\rho_N)=g(E)$, it follows from (\ref{B-lb-2+}) that
\begin{equation*}
 L_S(\rho_N\shs|\shs\varepsilon)\geq  g(E)-\varepsilon g(R_N(\varepsilon))-g(\varepsilon)\geq g(E)-\varepsilon g(N/\varepsilon)-g(\varepsilon),\quad \forall \varepsilon\in(0,(1+N)^{-1})
\end{equation*}
and
\begin{equation*}
 L_S(\rho_N\shs|\shs\varepsilon)\geq  g(E)-\varepsilon g(N/\varepsilon)-g(\varepsilon),\quad \forall \varepsilon\in[(1+N)^{-1},1].
\end{equation*}

Universal  lower bounds on $L_S(\rho\shs|\shs\varepsilon)$ can be obtained from Lemma \ref{b-lemma} in Section 3.\smallskip

\textbf{Proposition 1.}  \emph{Let $\rho$ be an arbitrary state in $\S(\H)$. Then
\begin{equation}\label{B-lb-3+}
 L_S(\rho\shs|\shs\varepsilon)\geq S([\rho-\varepsilon I_{\H}]_+)-g(\varepsilon)
\end{equation}
for any $\varepsilon\in(0,1]$, where $[\rho-\varepsilon I_{\H}]_+$ is the
positive part of the operator $\rho-\varepsilon I_{\H}$ and $S(\cdot)$ is the extended von Neumann entropy defined in (\ref{S-ext}).}

\emph{If $\rho$ is  a state with finite entropy then
\begin{equation}\label{B-lb-3++}
 L_S(\rho\shs|\shs\varepsilon)\geq S(\rho)-S(\rho\wedge\varepsilon I_{\H})-g(\varepsilon)
\end{equation}
for any $\varepsilon\in(0,1]$, where $\rho\wedge\varepsilon I_{\H}$ is the operator defined in (\ref{2-op}).}\smallskip

\emph{The lower bounds (\ref{B-lb-3+}) and (\ref{B-lb-3++}) are faithful: their r.h.s. tend to $\,S(\rho)\leq+\infty\,$ as $\,\varepsilon\to0$.}

\smallskip
\textbf{Remark 3.}
Since  $\rho=\rho\wedge\varepsilon I_{\H}+[\rho-\varepsilon I_{\H}]_+$, it follows from (\ref{w-k-ineq+}) that
$$
S(\rho)-S(\rho\wedge\varepsilon I_{\H})\geq S([\rho-\varepsilon I_{\H}]_+).
$$
So, lower bound (\ref{B-lb-3++}) is sharper than (\ref{B-lb-3+}). The advantage of lower bound (\ref{B-lb-3+})
consists in its validity for all states $\rho$ (including states with infinite entropy). Note also
that the r.h.s. of (\ref{B-lb-3+}) is easily calculated, since
\begin{equation*}
S([\rho-\varepsilon I_{\H}]_+)=\sum_{i\in I_\varepsilon} \eta(\lambda_i^{\rho}-\varepsilon)-\eta\!\left(\sum_{i\in I_\varepsilon}(\lambda_i^{\rho}-\varepsilon)\right),\quad \eta(x)=-x\ln x,
\end{equation*}
where $\{\lambda^{\rho}_i\}$ is the  sequences of eigenvalues of $\rho$ and $I_\varepsilon$ is the \emph{finite} set of all $i$ such that $\lambda_i^{\rho}>\varepsilon$.

For practical application of the lower bound (\ref{B-lb-3+}) it may be convenient to use the inequality
\begin{equation}\label{conv-ineq}
S([\rho-\varepsilon I_{\H}]_+)\geq\sum_{i\in I_\varepsilon} \eta(\lambda_i^{\rho})-\eta\!\left(\sum_{i\in I_\varepsilon}\lambda_i^{\rho}\right)-\varepsilon|I_{\varepsilon}|\ln|I_{\varepsilon}|-h_2(\varepsilon|I_{\varepsilon}|)
\end{equation}
valid for any operator $\rho$ in $\T_+(\H)$ such that $\Tr\rho\leq1$ and $\varepsilon\in (0,\Tr\rho)$,
where $|I_{\varepsilon}|$ denotes the cardinality of the set $I_\varepsilon$ of all $i$ such that $\lambda_i^{\rho}>\varepsilon$. Since
$$
[\rho-\varepsilon I_{\H}]_+=\sum_{i\in I_\varepsilon}\lambda_i^{\rho}|\varphi_i\rangle\langle\varphi_i|-\sum_{i\in I_\varepsilon}\varepsilon|\varphi_i\rangle\langle\varphi_i|,\quad
$$
inequality (\ref{conv-ineq})
follows from the right inequality in (\ref{w-k-ineq+}) and the inequality
$$
(x+y)h_2\left(\frac{x}{x+y}\right)\leq h_2(x)
$$
valid for any nonnegative numbers $x$ and $y$ such that $x+y\leq1$, which can be proved by noting that $\,\{x,1-x\}=(1-x-y)\{0,1\}+(x+y)\!\left\{x(x+y)^{-1},y(x+y)^{-1}\right\}\,$
and by using concavity of the Shannon entropy.
\smallskip

\emph{Proof of Proposition 1.} Let $\rho$ and  $\sigma$ be any states in $\S(\H)$ such that $\textstyle\frac{1}{2}\|\rho-\sigma\|_1\leq\varepsilon$.
Let $\{\lambda^{\rho}_i\}_{i=0}^{+\infty}$ and $\{\lambda^{\sigma}_i\}_{i=0}^{+\infty}$ be the  sequences of eigenvalues of these states  arranged in the non-increasing order and $\{\varphi_i\}_{i=0}^{+\infty}$ the basis in $\H$  such that
$\rho=\sum_{i=0}^{+\infty} \lambda^{\rho}_i |\varphi_i\rangle\langle\varphi_i|$.

Consider the state
\begin{equation*}
\hat{\sigma}=\sum_{i=0}^{+\infty} \lambda^{\sigma}_i |\varphi_i\rangle\langle\varphi_i|.
\end{equation*}
The Mirsky ineqiality (\ref{Mirsky-ineq+}) implies that
$$
\textstyle\frac{1}{2}\|\rho-\hat{\sigma}\|_1\leq\frac{1}{2}\|\rho-\sigma\|_1\leq \varepsilon.
$$
So, since the states $\rho$ and $\hat{\sigma}$ commute and the von Neumann entropy is a concave nonnegative function on $\S(\H)$  satisfying inequality (\ref{w-k-ineq})  and taking finite values  at finite rank states, Lemma \ref{b-lemma}B in Section 3  with $\S_0=\S(\H)$ implies that
$$
S(\sigma)=S(\hat{\sigma})\geq S([\rho-\varepsilon I_{\H}]_+)-g(\varepsilon),
$$
while  Lemma \ref{b-lemma}A implies that
$$
S(\sigma)=S(\hat{\sigma})\geq S(\rho)-S(\rho\wedge\varepsilon I_{\H})-g(\varepsilon)
$$
provided that $S(\rho)<+\infty$. $\Box$\medskip\pagebreak

Inequality (\ref{B-lb-3+}) means that
$$
S(\sigma)\geq S([\rho-\varepsilon I_{\H}]_+)-g(\varepsilon)
$$
for any states $\rho$ and  $\sigma$ in $\S(\H)$ such that  $\frac{1}{2}\|\rho-\sigma\|_1\leq\varepsilon$.

Inequality (\ref{B-lb-3++}) can be written as the one-side continuity bound
$$
S(\rho)-S(\sigma)\leq S(\rho\wedge\varepsilon I_{\H})+g(\varepsilon).
$$
valid for any states $\rho$ and  $\sigma$ in $\S(\H)$ such that  $\frac{1}{2}\|\rho-\sigma\|_1\leq\varepsilon$ and
$S(\rho)<+\infty$.

Analysis of concrete examples shows that the accuracy of the local lower bounds given by Proposition 1
is significantly lower than the accuracy of the lower bounds (\ref{B-lb-1}) and (\ref{B-lb-2+}) applicable in the cases $\,\rank\rho<+\infty\,$
and $\,S(\rho)<+\infty\,$ correspondingly (in accordance with Remark 2 at the end of Section 3).
The advantages of lower bounds (\ref{B-lb-3+}) and (\ref{B-lb-3++}) are their \emph{universality and simplicity}.
This is especially true for lower bound (\ref{B-lb-3+}) valid for all states in $\S(\H)$ (including states $\rho$ with $S(\rho)=+\infty$) and
easily computable.\smallskip

\textbf{Example 2.} Consider the state
$$
\rho_*=\sum_{k=5}^{+\infty} \frac{|k\rangle\langle k|}{ck\ln^2k},\quad \textrm{where}\quad c=\sum_{k=5}^{+\infty} \frac{1}{k\ln^2k}<1.
$$
It is easy to see that $S(\rho_*)=+\infty$. For given sufficiently small $\varepsilon\in(0,c]$ let $n_\varepsilon$ be the maximal $n\in\mathbb{N}\cap[5,+\infty)$ such that $n\ln^2 n\leq 1/\varepsilon$. Since $n_\varepsilon\ln^2 n_\varepsilon\leq 1/\varepsilon$, by applying inequality (\ref{conv-ineq}) we obtain
\begin{equation}\label{p-est}
\begin{array}{rl}
\displaystyle S([c\rho_*-\varepsilon I_{\H}]_+)\!\!& \displaystyle \geq\,\sum_{k=5}^{n_\varepsilon}\eta\left(\frac{1}{k\ln^2k}\right)-\eta\!\left(\sum_{k=5}^{n_\varepsilon}\frac{1}{k\ln^2k}\right)-\varepsilon n_{\varepsilon}\ln n_{\varepsilon}-h_2(\varepsilon n_{\varepsilon}) \\
& \displaystyle\geq\,\sum_{k=5}^{n_\varepsilon}\eta\!\left(\frac{1}{k\ln^2k}\right)-\frac{1}{e}-\frac{1}{\ln n_{\varepsilon}}-h_2\!\left(\frac{1}{\ln^2 n_{\varepsilon}}\right).
\end{array}
\end{equation}
We have
$$
\sum_{k=5}^{n_\varepsilon}\frac{\ln(k\ln^2k)}{k\ln^2k}\geq \int_5^{n_\varepsilon+1} \frac{dx}{x\ln x}=\ln  \frac{\ln(n_\varepsilon+1)}{\ln5}\geq\ln  \frac{-\ln\varepsilon}{3\ln5},
$$
where the obvious inequality $n_\varepsilon+1\geq \varepsilon^{-1/3}$ was used. Hence, by using (\ref{p-est}) we obtain
$$
S([\rho_*-\varepsilon I_{\H}]_+)=\frac{1}{c}\,S([c\rho_*-c\varepsilon I_{\H}]_+)\geq \frac{1}{c}\left(\ln  \frac{-\ln(c\varepsilon)}{3\ln5}-\frac{1}{e}-\frac{1}{\ln n_{c\varepsilon}}-h_2\!\left(\frac{1}{\ln^2 n_{c\varepsilon}}\right)\!\right)
$$
for all sufficiently small $\varepsilon>0$.  Thus, since $\,n_\varepsilon\to+\infty\,$ as $\,\varepsilon\to0$, lower bound (\ref{B-lb-3++}) implies that
$$
L_S(\rho_*|\shs\varepsilon)\geq\frac{1}{c}\ln  \frac{-\ln(c\varepsilon)}{3\ln5}-\frac{1}{ce}+o(1)=O(\ln(-\ln\varepsilon))\quad \textrm{as}\;\; \varepsilon\to0.
$$
This is a lower estimate on the \emph{unknown} rate of convergence of $L_S(\rho_*|\shs\varepsilon)$ to $S(\rho_*)=+\infty$ as $\varepsilon\to0$.

\subsection{Energy-type functionals and the quantum relative entropy}

Let $H$ be a positive operator on a separable Hilbert space $\H$ with representation (\ref{H-form}). Then the function
$$
E_{H}(\rho)=\Tr H\rho
$$
defined according to the rule (\ref{H-fun}) is an affine lower semicontinuous function  on $\S(\H)$ taking
values in $[0,+\infty]$. If $H$ is the Hamiltonian (energy observable) of some quantum system described by the space $\H$
then $E_H(\rho)$ is the mean energy of a state $\rho$.

The lower semicontinuity of $E_H$ implies that for any state $\rho$ in $\S(\H)$ the quantity
\begin{equation}\label{E-fun}
L_{E_H}(\rho\shs|\shs\varepsilon)=\inf\left\{E_H(\sigma)\,|\,\sigma\in\S(\H),\, \textstyle\frac{1}{2}\|\rho-\sigma\|_1\leq\varepsilon\right\}
\end{equation}
tends to $E_H(\rho)\leq+\infty\,$ as $\,\varepsilon\to0^+$. Lemma \ref{b-lemma} in Section 3 gives the following\smallskip

\textbf{Proposition 2.}   \emph{Let $H$ be a positive operator on $\H$ with representation (\ref{H-form}) and $\rho$ a state in $\S(\H)$ such that
$\supp\rho$ lies within the closure $\H_\mathcal{T}$ of the linear span of $\mathcal{T}\doteq\left\{\tau_k\right\}_{k=0}^{+\infty}$.\footnote{The support $\mathrm{supp}\rho$ of a state $\rho$ is the closed subspace spanned by the eigenvectors of $\rho$ corresponding to its positive eigenvalues.} Then
\begin{equation}\label{H-LB+}
 L_{E_H}(\rho\shs|\shs\varepsilon)\geq \sum_{k\in I_{\varepsilon}}E_k(\langle\tau_k|\rho|\tau_k\rangle-\varepsilon)\quad \forall\varepsilon\in(0,1],
\end{equation}
where $I_\varepsilon$ is the set of all $\,k$ such that $\langle\tau_k|\rho|\tau_k\rangle>\varepsilon$.}\smallskip

\emph{The lower bound (\ref{H-LB+}) is faithful: its r.h.s. tends to $\,E_H(\rho)\leq+\infty$ as $\,\varepsilon\to0$.}\smallskip

\textbf{Remark 4.}
The series in the r.h.s. of (\ref{H-LB+}) contains a finite number of nonzero summands and, hence, it is finite and easily calculated for any $\varepsilon>0$.

The condition $\supp\rho\subseteq\H_\mathcal{T}$ in Proposition 2 is essential. If $\supp\rho\nsubseteq\H_\mathcal{T}$
then it is easy to see that $L_{E_H}(\rho\shs|\shs\varepsilon)=E_H(\rho)=+\infty\,$ for all sufficiently small $\,\varepsilon>0$.\smallskip

\emph{Proof.} Let $\sigma$ be any state in $\S(\H)$ such that $\textstyle\frac{1}{2}\|\rho-\sigma\|_1\leq\varepsilon$.

Assume that  $\,E_H(\sigma)<+\infty$. This implies that $\supp\sigma\subseteq\H_\mathcal{T}$. Since the same
inclusion holds for the state $\rho$ by the condition, the operators
\begin{equation*}
\hat{\rho}=\sum_{k=0}^{+\infty} \langle\tau_k|\rho|\tau_k\rangle|\tau_k\rangle\langle\tau_k|\quad \textrm{and}\quad
\hat{\sigma}=\sum_{k=0}^{+\infty} \langle\tau_k|\sigma|\tau_k\rangle|\tau_k\rangle\langle\tau_k|
\end{equation*}
are states in $\S(\H)$ such that $E_H(\hat{\rho})=E_H(\rho)$ and $E_H(\hat{\sigma})=E_H(\sigma)$.

Since $\hat{\rho}=\Pi(\rho)$ and $\hat{\sigma}=\Pi(\sigma)$, where $\Pi$ is the  channel $\,\varrho\mapsto \sum_{k=0}^{+\infty} \langle\tau_k|\varrho|\tau_k\rangle|\tau_k\rangle\langle\tau_k|\,$ from $\T(\H_\mathcal{T})$ to itself, we have
$$
\textstyle\frac{1}{2}\|\hat{\rho}-\hat{\sigma}\|_1\leq\frac{1}{2}\|\rho-\sigma\|_1\leq \varepsilon.
$$
So, since the states $\hat{\rho}$ and $\hat{\sigma}$ commute and $E_H$ is an affine nonnegative function on $\S(\H_\mathcal{T})$
taking finite values at finite rank states in $\S(\H_\mathcal{T})$ diagonizable in the basis $\{\tau_k\}$, Lemma \ref{b-lemma}B in Section 3
with $\S_0=\S(\H_\mathcal{T})$ implies that
$$
E_H(\sigma)=E_H(\hat{\sigma})\geq E_H([\hat{\rho}-\varepsilon I_{\H}]_+)=\sum_{k\in I_{\varepsilon}}E_k(\langle\tau_k|\rho|\tau_k\rangle-\varepsilon).
$$
To complete the proof of (\ref{H-LB+}) it suffices to note that the above inequality holds trivially if $\,E_H(\sigma)=+\infty$.

The last claim of the proposition is easily verified. $\Box$\smallskip

\textbf{Example 3.} Despite the fact that the lower bound (\ref{H-LB+}) is not too accurate in general, there exist
states $\rho$ and $\varepsilon>0$ for which an equality holds in (\ref{H-LB+}). Assume that the operator $H$ satisfies condition (\ref{star}) and $\rho=|\tau_k\rangle\langle\tau_k|$, $k>0$. Then it is easy to see that the infimum in (\ref{E-fun}) is attained at the state $\,\sigma=\varepsilon|\tau_0\rangle\langle\tau_0|+(1-\varepsilon)\rho\,$ for any $\varepsilon\in(0,1]$ and hence both sides of (\ref{H-LB+}) are equal to $E_k(1-\varepsilon)$.

\medskip

The \emph{quantum relative entropy} for two states $\rho$ and
$\sigma$ in $\mathfrak{S}(\mathcal{H})$ is defined as
\begin{equation}\label{qre-def}
D(\rho\,\|\shs\sigma)=\sum_i\langle
\varphi_i|\,\rho\ln\rho-\rho\ln\sigma\,|\varphi_i\rangle,
\end{equation}
where $\{\varphi_i\}$ is the orthonormal basis of
eigenvectors of the state $\rho$ and it is assumed that
$D(\rho\,\|\sigma)=+\infty$ if $\,\mathrm{supp}\rho\shs$ is not
contained in $\shs\mathrm{supp}\shs\sigma$ \cite{H-SCI,L-2,W}.

Since the function $D_{\omega}(\rho)\doteq D(\rho\shs\|\shs\omega)$ is lower semicontinuous  on $\S(\H)$ for any given $\omega\in\S(\H)$, the quantity
\begin{equation*}
L_{D_{\omega}}(\rho\shs|\shs\varepsilon)=\inf\left\{D(\sigma\|\shs\omega)\,|\,\sigma\in\S(\H),\, \textstyle\frac{1}{2}\|\rho-\sigma\|_1\leq\varepsilon\right\}
\end{equation*}
tends to $D(\rho\|\shs\omega)\leq+\infty\,$ as $\,\varepsilon\to0$ for any $\rho\in\S(\H)$.

Our technique
allows to obtain lower bounds on $L_{D_{\omega}}(\rho\shs|\shs\varepsilon)$ which are faithful only if either
$\,\rank\omega<+\infty\,$ or $\,[\rho,\omega]=0$.\footnote{In the case $\,\rank\omega<+\infty\,$ faithful lower bounds on $L_{D_{\omega}}(\rho\shs|\shs\varepsilon)$
can be also derived from the continuity bounds for the function $\,\rho\mapsto D(\rho\shs\|\shs\omega)\,$ in the finite-dimensional settings
obtained in \cite[Section 5.8]{Hanson} and in \cite[Section 5.2.4]{Capel}.} It also gives
faithful lower bounds on the  quantities
\begin{equation}\label{RE-fun+}
L^{d}_{D_{\omega}}(\rho\shs|\shs\varepsilon)=\inf\left\{D(\sigma\|\shs\omega)\,|\,\sigma\in\S(\H),\; \rank\sigma\leq d,\; \textstyle\frac{1}{2}\|\rho-\sigma\|_1\leq\varepsilon\right\},
\end{equation}
where it is assumed that $\inf\emptyset=+\infty$, and
\begin{equation}\label{RE-fun++}
L^{\rm com}_{D_{\omega}}(\rho\shs|\shs\varepsilon)=\inf\left\{D(\sigma\|\shs\omega)\,|\,\sigma\in\S(\H),\; [\rho,\sigma]=0, \; \textstyle\frac{1}{2}\|\rho-\sigma\|_1\leq\varepsilon\right\}.
\end{equation}
Note that we do not impose any constraints on the state $\rho$ in (\ref{RE-fun+}) and (\ref{RE-fun++}).

The quantities $L^{d}_{D_{\omega}}(\rho\shs|\shs\varepsilon)$ and $L^{\rm com}_{D_{\omega}}(\rho\shs|\shs\varepsilon)$ are different upper bounds on $L_{D_{\omega}}(\rho\shs|\shs\varepsilon)$. If $\omega$ is a finite rank state then it is easy to see that $L^d_{D_{\omega}}(\rho\shs|\shs\varepsilon)=L_{D_{\omega}}(\rho\shs|\shs\varepsilon)$ for all $d\geq \rank\omega$.

The lower semicontinuity of $D_{\omega}$ implies that  $L^{\rm com}_{D_{\omega}}(\rho\shs|\shs\varepsilon)$ tends to $D(\rho\shs\|\shs\omega)\leq+\infty\,$ as $\,\varepsilon\to0$. The same is true for $L^{d}_{D_{\omega}}(\rho\shs|\shs\varepsilon)$  provided that $\rank\rho\leq d$. \pagebreak

\textbf{Proposition 3.}  \emph{Let $\omega$ be an arbitrary state with the spectral decomposition $\omega=\sum_{k=0}^{+\infty} \lambda^{\omega}_k|\psi_k\rangle\langle\psi_k|$ and $\rho$  a state such that $\,\supp\rho\subseteq\supp\omega$.}\smallskip

\noindent A) \emph{If $\rho=\sum_{k=0}^{+\infty} \lambda^{\rho}_k|\varphi_k\rangle\langle\varphi_k|$ is a spectral decomposition\footnote{We assume that $\{\varphi_k\}$ is basis in $\H$, so some $\lambda^{\rho}_k$ in this decomposition may be equal to zero.} of $\rho$ then
\begin{equation}\label{RE-LB+D}
L_{D_{\omega}}(\rho\shs|\shs\varepsilon)\geq \sum_{k\in I_{\varepsilon}}(\lambda^{\rho}_k-\varepsilon)\left(\ln(\lambda^{\rho}_k-\varepsilon)-\ln\langle\varphi_k|\omega|\varphi_k\rangle\right)-g(\varepsilon)-
\tilde{h}_2(\varepsilon)-\eta(1-r_{\varepsilon})
\end{equation}
for any $\varepsilon\in(0,1]$, where $I_\varepsilon$ is the set of all $\,k$ such that $\lambda^{\rho}_k>\varepsilon$, $\tilde{h}_2$ and $g$ are the functions defined in (\ref{h+}) and (\ref{g-def}), $\,\eta(x)=-x\ln x\,$ and $\,r_{\varepsilon}\doteq\Tr[\rho-\varepsilon I_{\H}]_+=\sum_{k\in I_{\varepsilon}}(\lambda^{\rho}_k-\varepsilon)$.}
\smallskip

\noindent B) \emph{If $\,S(\rho)<+\infty$ and $\,d\geq2\,$ is a natural number then
\begin{equation}\label{RE-LB+A}
L^{d}_{D_{\omega}}(\rho\shs|\shs\varepsilon)\geq \sum_{k\in J_{\varepsilon}}(-\ln\lambda^{\omega}_k)(\langle\psi_k|\rho|\psi_k\rangle-\varepsilon)-S(\rho)-\varepsilon\ln(d-1)-h_2(\varepsilon)
\end{equation}
for any $\,\varepsilon\in(0,1-1/d]$, where $J_\varepsilon$ is the set of all $k$ such that $\langle\psi_k|\rho|\psi_k\rangle>\varepsilon$.}
\smallskip

\noindent C) \emph{If either $D(\rho\shs\|\shs\omega)<+\infty\,$ or $ \,-\langle\varphi|\ln\omega|\varphi\rangle<+\infty\,$ for any eigenvector $\varphi$ of the state $\rho$ corresponding to its nonzero eigenvalue then
\begin{equation}\label{RE-LB+C}
\!L^{\rm com}_{D_{\omega}}(\rho\shs|\shs\varepsilon)\geq r_{\varepsilon}D(r^{-1}_{\varepsilon}[\rho-\varepsilon I_{\H}]_+\|\shs\omega)-g(\varepsilon)-\tilde{h}_2(\varepsilon)-h_2(1-r_{\varepsilon}),\quad r_{\varepsilon}\doteq\Tr[\rho-\varepsilon I_{\H}]_+,
\end{equation}
for any $\varepsilon\in(0,1]$, where $\tilde{h}_2$ and $g$ are the functions defined in (\ref{h+}) and (\ref{g-def}).}
\smallskip

\emph{The lower bounds (\ref{RE-LB+A}) and (\ref{RE-LB+C}) are faithful: their r.h.s. tend to $\,D(\rho\shs\|\shs\omega)\leq+\infty$ as $\,\varepsilon\to0$.}
\smallskip

\emph{The lower bound (\ref{RE-LB+D}) is faithful if $\,[\rho,\omega]=0$ and the state $\omega$ is diagonizable in the basis $\{\varphi_k\}$. In the general case its right hand side tends to $\,D(\rho\shs\|\shs\widetilde{\omega})\leq D(\rho\shs\|\shs\omega)\leq+\infty$ as $\,\varepsilon\to0$, where $\,\widetilde{\omega}=\sum_{k=0}^{+\infty} \langle\varphi_k|\omega|\varphi_k\rangle|\varphi_k\rangle\langle\varphi_k|$.}\medskip

\textbf{Remark 5.}
The series in the r.h.s. of (\ref{RE-LB+D}) and (\ref{RE-LB+A}) contain a finite number of nonzero summands and, hence, they are finite and easily calculated for any $\varepsilon>0$. In the proof Proposition 3 it is shown that the condition in part C implies
that the r.h.s. of (\ref{RE-LB+C}) is finite and determined by a finite sum for any $\varepsilon>0$.\smallskip

The condition $\supp\rho\subseteq\supp\omega$ in Proposition 3 is essential. If $\supp\rho\nsubseteq\supp\omega$
then it is easy to see that $\,L_{D_{\omega}}(\rho\shs|\shs\varepsilon)=L^d_{D_{\omega}}(\rho\shs|\shs\varepsilon)=L^{\rm com}_{D_{\omega}}(\rho\shs|\shs\varepsilon)=+\infty\,$ for all sufficiently small $\varepsilon>0$.
\smallskip

\emph{Proof.} A) Let $\sigma$ be a state such that $\textstyle\frac{1}{2}\|\rho-\sigma\|_1\leq\varepsilon$.  Consider the channel
$$
\Pi(\varrho)=\sum_{k=0}^{+\infty} \langle\varphi_k|\varrho|\varphi_k\rangle|\varphi_k\rangle\langle\varphi_k|,\quad \varrho\in\S(\H).
$$
By monotonicity of the relative entropy  we have
\begin{equation}\label{D-1}
 D(\sigma\|\shs\omega)\geq D(\tilde{\sigma}\|\shs\tilde{\omega})
\end{equation}
where $\tilde{\sigma}=\Pi(\sigma)$ and $\tilde{\omega}=\Pi(\omega)$ are states in $\S(\H)$.

The function $\,D_{\tilde{\omega}}(\varrho)=D(\varrho\shs\|\shs\tilde{\omega})\,$ is nonnegative convex and satisfies inequality (\ref{LAA-1}) with $a_f=h_2$ \cite[Proposition 5.24]{O&P}. Since the states $\rho$ and $\tilde{\omega}$ are diagonizable in the basis $\{\varphi_k\}$ and $\,\supp\rho\subseteq\supp\tilde{\omega}\,$ by the condition $\,\supp\rho\subseteq\supp\omega$,
we have
$$
r_{\varepsilon}D(r^{-1}_{\varepsilon}[\rho-\varepsilon I_{\H}]_+\|\shs\tilde{\omega})=\sum_{k\in I_{\varepsilon}}(\lambda^{\rho}_k-\varepsilon)\left(\ln(\lambda^{\rho}_k-\varepsilon)-\ln\langle\varphi_k|\omega|\varphi_k\rangle\right)+\eta(r_{\varepsilon})<+\infty
$$
for any $\varepsilon>0$. So, as the states $\rho$ and $\tilde{\sigma}$ commute and
$$
\textstyle\frac{1}{2}\|\rho-\tilde{\sigma}\|_1=\textstyle\frac{1}{2}\|\Pi(\rho)-\Pi(\sigma)\|_1\leq\textstyle\frac{1}{2}\|\rho-\sigma\|_1\leq\varepsilon
$$
by the monotonicity of the trace norm  under action of a channel, it follows from  Lemma \ref{b-lemma}B in Section 3 with $\,\S_0=\left\{\varrho=\sum_{k=0}^{+\infty} p_k|\varphi_k\rangle\langle\varphi_k|\;|\, \{p_k\}\in\P_1\right\}\,$ that
$$
D(\tilde{\sigma}\|\shs\tilde{\omega})\geq r_{\varepsilon}D(r^{-1}_{\varepsilon}[\rho-\varepsilon I_{\H}]_+\|\shs\tilde{\omega})-g(\varepsilon)-
\tilde{h}_2(\varepsilon)-h_2(1-r_{\varepsilon})
$$
This inequality and (\ref{D-1}) implies (\ref{RE-LB+D}).\smallskip

B) Assume that $\sigma$ is  a state in $\S(\H)$ such that $\rank\sigma\leq d$ and $\textstyle\frac{1}{2}\|\rho-\sigma\|_1\leq\varepsilon\leq1-1/d$.
Then $S(\sigma)<+\infty$ and hence
$$
D(\sigma\|\shs\omega)=\Tr\sigma(-\ln\omega)-S(\sigma)=\Tr\sigma(-\ln\omega)-S(\rho)+[S(\rho)-S(\sigma)].
$$
By Proposition 2 in \cite{LCB}  (proved by using  Audenaert's continuity bound \cite{Aud}) we have
\begin{equation}\label{t-ineq}
S(\rho)-S(\sigma)\geq -\varepsilon\ln(d-1)-h_2(\varepsilon)
\end{equation}
(regardless of the rank of $\rho$). Thus, Proposition 2 implies that
$$
D(\sigma\|\shs\omega)\geq\sum_{k\in J_\varepsilon}(-\ln\lambda^{\omega}_k)(\langle\psi_k|\rho|\psi_k\rangle-\varepsilon)-S(\rho)-\varepsilon\ln(d-1)-h_2(\varepsilon).
$$

C) The function $\,D_{\omega}(\varrho)=D(\varrho\shs\|\shs\omega)\,$ is nonnegative convex and satisfies inequality (\ref{LAA-1}) with $a_f=h_2$ \cite[Proposition 5.24]{O&P}. The condition in part C implies that $\,D(r^{-1}_{\varepsilon}[\rho-\varepsilon I_{\H}]_+\|\shs\omega)<+\infty$ for all $\varepsilon\in(0,1]$. Indeed, if $\,D(\rho\|\shs\omega)<+\infty$ then this follows from the validity of inequality (\ref{LAA-1})
for the function $D_{\omega}$, since $[\rho-\varepsilon I_{\H}]_+\leq \rho$. If $ \,-\langle\varphi|\ln\omega|\varphi\rangle<+\infty\,$ then
this follows from the convexity of the function $D_{\omega}$. Thus, Lemma \ref{b-lemma}B  with $\S_0=\S(\H)$ shows that
$$
D_{\omega}(\sigma)\geq r_{\varepsilon}D_{\omega}(r^{-1}_{\varepsilon}[\rho-\varepsilon I_{\H}]_+)-g(\varepsilon)-\tilde{h}_2(\varepsilon)-h_2(1-r_{\varepsilon})
$$
for any state $\sigma$ in $\S(\H)$ such that $[\rho,\sigma]=0$ and $\textstyle\frac{1}{2}\|\rho-\sigma\|_1\leq\varepsilon\leq1$. This inequality
implies (\ref{RE-LB+C}).

To prove the last claims of the proposition it suffices to note that
$$
\lim_{\varepsilon\to0^+}\sum_{k\in I_{\varepsilon}}(\lambda^{\rho}_k-\varepsilon)\left(\ln(\lambda^{\rho}_k-\varepsilon)-\ln\langle\varphi_k|\omega|\varphi_k\rangle\right)=D(\rho\shs\|\shs\tilde{\omega})\leq+\infty,
$$
$$
\lim_{\varepsilon\to0^+}\sum_{k\in J_{\varepsilon}}(\langle\psi_k|\rho|\psi_k\rangle-\varepsilon)(-\ln\lambda^{\omega}_k)=-\Tr\rho\ln\omega\leq+\infty
$$
and that
$$
\lim_{\varepsilon\to0^+}r_{\varepsilon}D(r^{-1}_{\varepsilon}[\rho-\varepsilon I_{\H}]_+\|\shs\omega)=D(\rho\shs\|\shs\omega)\leq+\infty.
$$
The first two limit relations are obvious, the last one can be easily shown by using the lower semicontinuity of the nonnegative function $\,D_{\omega}(\rho)=D(\rho\shs\|\shs\omega)\,$
and  the validity of  inequality (\ref{LAA-1}) for this function with $a_f=h_2$ mentioned before. $\Box$\smallskip

By using the arguments from the proof of Proposition 3B and Proposition 1 in \cite{LCB} one can obtain the following\smallskip

\textbf{Proposition 4.} \emph{Let $\omega$ be an arbitrary state in $\S(\H)$  with the spectral decomposition $\omega=\sum_{k=0}^{+\infty} \lambda^{\omega}_k|\psi_k\rangle\langle\psi_k|$ and $\rho$  a state such that $\supp\rho\subseteq\supp\omega$. If $H$ is a positive operator on $\H$ satisfying conditions (\ref{H-cond}) and (\ref{star}) and $E>0$ then
\begin{equation}\label{RE-LB+B}
D(\sigma\shs\|\shs\omega)\geq \sum_{k\in J_{\varepsilon}}(-\ln\lambda^{\omega}_k)(\langle\psi_k|\rho|\psi_k\rangle-\varepsilon)-S(\rho)-\varepsilon F_H(E/\varepsilon)-g(\varepsilon)
\end{equation}
for any state $\sigma$ in $\S(\H)$ such that  $\textstyle\frac{1}{2}\|\rho-\sigma\|_1\leq\varepsilon\leq1$ and $\Tr H\sigma\leq E$,
where $J_\varepsilon$ is the set of all $k$ such that $\langle\psi_k|\rho|\psi_k\rangle>\varepsilon$, $F_H$ and $g$ are the functions defined in (\ref{F-def}) and (\ref{g-def}).}

\emph{The right hand side of (\ref{RE-LB+B}) tends to $D(\rho\shs\|\shs\omega)\leq+\infty\,$ as $\,\varepsilon\to0$.}

\smallskip

\emph{Proof.} To prove (\ref{RE-LB+B}) it suffices to note that the condition $\Tr H\sigma\leq E$ implies the finiteness of $S(\sigma)$
and to replace inequality (\ref{t-ineq}) in the proof of Proposition 3B by the inequality
$$
S(\rho)-S(\sigma)\geq -\varepsilon F_H(E/\varepsilon)-g(\varepsilon),
$$
which follows from Proposition 1 in \cite{LCB}.

The last claim of the proposition follows from the equivalence of (\ref{H-cond}) and (\ref{H-cond-a}). $\Box$\smallskip

\textbf{Note:} In Proposition 4 no constrains on the state $\rho$ are imposed. So, the set of states
$\sigma$ such that  $\textstyle\frac{1}{2}\|\rho-\sigma\|_1\leq\varepsilon\leq1$ and $\Tr H\sigma\leq E$ may be empty for some $E$ and $\varepsilon$.

\subsection{Quantum conditional entropy of quantum-classical states}

The \emph{quantum conditional entropy} (QCE) of a state $\rho$ of a finite-dimensional bipartite system $AB$ is defined as
\begin{equation}\label{ce-def}
S(A|B)_{\rho}=S(\rho)-S(\rho_{B}).
\end{equation}
Definition (\ref{ce-def}) remains valid for a state $\rho$ of an infinite-dimensional bipartite system $AB$
with finite marginal entropies
$S(\rho_A)$ and $S(\rho_B)$  (since the finiteness of $S(\rho_A)$ and $S(\rho_B)$ are equivalent to the finiteness of $S(\rho)$ and $S(\rho_B)$).
For a state $\rho$ with finite $S(\rho_A)$ and arbitrary $S(\rho_B)$ one can define the QCE
by the formula
\begin{equation}\label{ce-ext}
S(A|B)_{\rho}=S(\rho_{A})-D(\rho\shs\Vert\shs\rho_{A}\otimes\rho_{B})
\end{equation}
proposed and analysed by Kuznetsova in \cite{Kuz}, where $D(\cdot\shs\Vert\shs\cdot)$ is the quantum relative entropy defined in (\ref{qre-def}) (the finiteness of $S(\rho_{A})$ implies the finiteness of $D(\rho\shs\Vert\shs\rho_{A}\otimes\rho_{B})$). The QCE  extented by formula (\ref{ce-ext}) to the convex set $\,\{\rho\in\S(\H_{AB})\,|\,S(\rho_A)<+\infty\}\,$ possesses all basic properties of the QCE valid in finite dimensions \cite{Kuz}.

In this subsection we consider local lower bounds on the (extended) quantum conditional entropy defined in (\ref{ce-ext}) restricting attention to
quantum-classical (q-c) states, i.e. states $\rho$ in $\S(\H_{AB})$ having the form
\begin{equation}\label{qc-states}
\rho=\sum_{k} p_k \rho_k\otimes |k\rangle\langle k|,
\end{equation}
where $\{p_k,\rho_k\}$ is an ensembles of states in $\S(\H_A)$ and $\{|k\rangle\}$ a fixed orthonormal basis in $\H_B$.\footnote{An ensemble $\{p_k,\rho_k\}$ is a collection $\{\rho_k\}$ of quantum states with a probability distribution $\{p_k\}$.}
By using definition (\ref{ce-ext}) of QCE one can show (see the proof of Corollary 3 in \cite{Wilde-CB}) that
\begin{equation}\label{ce-rep}
S(A|B)_{\rho}=\sum_kp_kS(\rho_k).
\end{equation}
This expression allows us to define the QCE on the set $\S_{\mathrm{qc}}$ of all q-c states with representation (\ref{qc-states})
(including the q-c states $\rho$ with $\,S(\rho_A)=+\infty\,$ for which definition (\ref{ce-ext}) does not work).
It  means that the value of QCE at any q-c state $\rho$ determined by an ensemble $\{p_k,\rho_k\}$  coincides with the average entropy
of this ensemble. The properties of the von Neumann entropy imply that the QCE defined by expression (\ref{ce-rep}) on the convex set $\S_{\mathrm{cq}}$ satisfies
inequalities (\ref{LAA-1}) and (\ref{LAA-2}) with $a_f=0$ and $b_f=h_2$ with possible values $+\infty$ in both sides.

Since for any q-c states $\rho$ and $\sigma$ determined, respectively, by ensembles $\{p_k,\rho_k\}$ and $\{q_k,\sigma_k\}$ we have
$$
\|\rho-\sigma\|_1=\sum_k\|p_k\rho_k-q_k\sigma_k\|_1,
$$
expression (\ref{ce-rep}) shows, due to the lower semicontinuity of the von Neumann entropy, that the function $\rho\mapsto S(A|B)_{\rho}$ is lower semicontinuous on the set $\S_{\mathrm{cq}}$. So, the quantity
\begin{equation*}
L^{\rm qc}_{S(A|B)}(\rho\shs|\shs\varepsilon)=\inf\left\{S(A|B)_{\sigma}\,|\,\sigma\in\S_{\mathrm{qc}},\, \textstyle\frac{1}{2}\|\rho-\sigma\|_1\leq\varepsilon\right\}
\end{equation*}
tends to $S(A|B)_{\rho}\leq+\infty\,$ as $\,\varepsilon\to0$ for any q-c state $\rho$ in $\S_{\mathrm{cq}}$.\smallskip

The results of Section 4.2 in \cite{LCB} imply
easily computable lower bounds on $L^{\rm qc}_{S(A|B)}(\rho\shs|\shs\varepsilon)$.

Proposition 3A in \cite{LCB}  shows that
\begin{equation}\label{CE-LB-1}
L^{\rm qc}_{S(A|B)}(\rho\shs|\shs\varepsilon)\geq S(A|B)_{\rho}-\varepsilon\ln(\rank\rho_A)-g(\varepsilon)
\end{equation}
for any q-c state $\rho$ in $\S_{\mathrm{qc}}$ such that $\rank\rho_A$ is finite and arbitrary $\varepsilon\in(0,1]$, where $g$ is the function defined in (\ref{g-def}).\pagebreak

Proposition 3B in \cite{LCB}  shows that
\begin{equation}\label{CE-LB-2}
\!L^{\rm qc}_{S(A|B)}(\rho\shs|\shs\varepsilon)\geq S(A|B)_{\rho}-\varepsilon F_H(E^{\rm qc}_{H,\shs\varepsilon}(\rho))/\varepsilon)-g(\varepsilon)\geq S(A|B)_{\rho}-\varepsilon F_H(E/\varepsilon)-g(\varepsilon)
\end{equation}
for any q-c state $\rho$ in $\S_{\mathrm{qc}}$ such that $\Tr H\rho_A=E<+\infty$  and arbitrary $\varepsilon\in(0,1]$, where $H$ is  any positive operator on $\H_A$  satisfying conditions (\ref{H-cond}) and (\ref{star}), $F_H$ is the function defined in (\ref{F-def})
and\footnote{$(p_k\rho_k)\wedge\varepsilon I_{A}$  and $[p_k\rho_k-\varepsilon  I_{A}]_+$ are the operators defined in (\ref{2-op}).}
\begin{equation*}
E^{\rm qc}_{H,\shs\varepsilon}(\rho)\doteq \sum_{k}\Tr H((p_k\rho_k)\wedge\varepsilon I_{A})=E-\sum_{k}\Tr H[p_k\rho_k-\varepsilon I_{A}]_+=o(1)\quad\textrm{as }\;\;\varepsilon\to0.
\end{equation*}
The lower bound (\ref{CE-LB-2}) is applicable
to any q-c state  $\rho$  with finite $S(\rho_A)$, since for any such state there is a positive operator $H$ on $\H_A$ satisfying conditions (\ref{H-cond}) and (\ref{star}) such that $\Tr H\rho_A<+\infty$ \cite[Proposition 4]{EC}. As in the von Neumann entropy case (see Section 4.1), the operator $H$ can be treated as a "free parameter", with which one can optimize lower bound (\ref{CE-LB-2}).

Since one-side continuity bounds given by parts A and B of Proposition 3 in \cite{LCB} are asymptotically tight for large $\rank\rho_A$ and $\Tr H\rho_A$ correspondingly,
the lower bounds  (\ref{CE-LB-1}) and (\ref{CE-LB-2}) are quite accurate. They are faithful: the right hand sides of (\ref{CE-LB-1}) and (\ref{CE-LB-2})
tends to $S(A|B)_{\rho}\,$ as $\,\varepsilon\to0$  (for the last one this follows from the equivalence of  (\ref{H-cond}) and (\ref{H-cond-a})).

Coarse but universal and easily computable lower bounds on $\,L^{\rm qc}_{S(A|B)}(\rho\shs|\shs\varepsilon)\,$ can be obtained by using Lemma \ref{b-lemma} in Section 3. In the following proposition we assume that $S(A|B)$ is the homogeneous extension of the QCE to the cone generated by the set $\S_{\rm qc}$
of q-c states which is defined according to the rule (\ref{G-ext}). We use the notation in (\ref{2-op}).\smallskip

\textbf{Proposition 5.}  \emph{Let $\rho$ be an arbitrary q-c state in $\S_{\rm qc}$. Then
\begin{equation}\label{CE-LB-3+}
L^{\rm qc}_{S(A|B)}(\rho\shs|\shs\varepsilon)\geq S(A|B)_{\rho\ominus\varepsilon}-g(\varepsilon)\quad\forall\varepsilon\in(0,1],
\end{equation}
where}
$$
\rho\ominus\varepsilon\doteq\sum_k [p_k\rho_k-\varepsilon I_{A}]_+\otimes|k\rangle\langle k|.
$$

\emph{If $\,S(A|B)_{\rho}<+\infty\,$ then
\begin{equation}\label{CE-LB-3++}
 L^{\rm qc}_{S(A|B)}(\rho\shs|\shs\varepsilon)\geq S(A|B)_{\rho}-S(A|B)_{\rho\wedge\varepsilon}-g(\varepsilon)\quad\forall\varepsilon\in(0,1],
\end{equation}
where}
$$
\rho\wedge\varepsilon\doteq\sum_k \left((p_k\rho_k)\wedge\varepsilon I_{A}\right)\otimes|k\rangle\langle k|.
$$

\emph{The lower bounds (\ref{CE-LB-3+}) and (\ref{CE-LB-3++}) are faithful: their r.h.s. tend to $\,S(A|B)_{\rho}\leq+\infty$ as $\,\varepsilon\to0$.}\smallskip

Using expression (\ref{ce-rep}) one can rewrite Proposition 5 as follows.\smallskip

\begin{corollary}\label{CE-LB-3+c}  \emph{Let $\{p_k,\rho_k\}$ be an arbitrary ensemble of
states in $\S(\H)$. Then
\begin{equation}\label{CE-LB-3+c+}
\sum_kq_kS(\sigma_k)\geq \sum_kS([p_k\rho_k-\varepsilon I_{\H}]_+)-g(\varepsilon)
\end{equation}
for any ensemble $\{q_k,\sigma_k\}$ of states in $\S(\H)$ such that  $\frac{1}{2}\sum_k\|p_k\rho_k-q_k\sigma_k\|_1\leq\varepsilon$,
where $S(\cdot)$ in the r.h.s. is the extended von Neumann  entropy defined in (\ref{S-ext}).}\smallskip

\emph{If  $\,\sum_kp_kS(\rho_k)<+\infty\,$ then
\begin{equation}\label{CE-LB-3++c}
\sum_kq_kS(\sigma_k)\geq \sum_kp_kS(\rho_k)-\sum_kS((p_k\rho_k)\wedge\varepsilon I_{\H})-g(\varepsilon)
\end{equation}
for any ensemble $\{q_k,\sigma_k\}$ of states in $\S(\H)$ such that  $\frac{1}{2}\sum_k\|p_k\rho_k-q_k\sigma_k\|_1\leq\varepsilon$.}
\end{corollary}\medskip

\textbf{Remark 6.}
Since  $\,p_k\rho_k=(p_k\rho_k)\wedge\varepsilon I_{\H}+[p_k\rho_k-\varepsilon I_{\H}]_+$, it follows from (\ref{w-k-ineq+}) that
$$
\sum_kp_kS(\rho_k)-\sum_kS((p_k\rho_k)\wedge\varepsilon I_{\H})\geq \sum_kS([p_k\rho_k-\varepsilon I_{\H}]_+).
$$
So, lower bound (\ref{CE-LB-3++c}) is sharper than (\ref{CE-LB-3+c+}). The advantage of lower bound (\ref{CE-LB-3+c+})
consists in its validity for all ensembles $\{p_k,\rho_k\}$ (including ensembles with $\sum_kp_kS(\rho_k)=+\infty$). Note also
that the r.h.s. of (\ref{CE-LB-3+c+}) is easily calculated, since the series in the r.h.s. of (\ref{CE-LB-3+c+}) contains a finite number of nonzero summands and
all the operators  $[p_k\rho_k-\varepsilon I_{\H}]_+$ have finite rank.\smallskip

\emph{Proof of Proposition 5.} Let
$$
\rho=\sum_{k} p_k \rho_k\otimes |k\rangle\langle k|\quad \textrm{and}\quad\sigma=\sum_{k} q_k \sigma_k\otimes |k\rangle\langle k|
$$
be q-c states in $\S_{\rm qc}$  such that
$$
\|\rho-\sigma\|_1=\sum_k\|p_k\rho_k-q_k\sigma_k\|_1\leq 2\varepsilon.
$$
For each $k$ let $\{\varphi^k_i\}_i$ be an orthonormal basis in $\H_A$ such that
$$
\rho_k=\sum_{i} \lambda^{\rho_k}_i |\varphi^k_i\rangle\langle \varphi^k_i|,
$$
where $\{\lambda^{\rho_k}_i\}_i$ is a non-increasing sequence of eigenvalues of $\rho_k$.
Consider the q-c state
\begin{equation*}
\tilde{\sigma}=\sum_{k} q_k\tilde{\sigma}_k\otimes |k\rangle\langle k|,\quad\tilde{\sigma}_k=\sum_{i} \lambda^{\sigma_k}_i |\varphi^k_i\rangle\langle \varphi^k_i|,
\end{equation*}
where $\{\lambda^{\sigma_k}_i\}_i$ is a non-increasing sequence of eigenvalues of $\sigma_k$ for each $k$.

By Mirsky inequality (\ref{Mirsky-ineq+}) we have
$$
\|\rho-\tilde{\sigma}\|_1=\sum_k\|p_k\rho_k-q_k\tilde{\sigma}_k\|_1\leq\sum_k\|p_k\rho_k-q_k\sigma_k\|_1\leq 2\varepsilon.
$$

The representation (\ref{ce-rep}) implies that
$$
S(A|B)_{\sigma}=\sum_kq_kS(\sigma_k)=\sum_kq_kS(\tilde{\sigma}_k)=S(A|B)_{\tilde{\sigma}}.
$$

It is easy to see that any q-c state $\rho$ in $\S_{\rm qc}$
satisfies conditions (\ref{S-prop+}) and (\ref{S-prop++}) with $\S_0=\S_{\rm qc}$  and that
$$
\rho\wedge\varepsilon I_{AB}=\sum_k ((p_k\rho_k)\wedge\varepsilon I_{A})\otimes |k\rangle\langle k|, \quad[\rho-\varepsilon I_{AB}]_+=\sum_k [p_k\rho_k-\varepsilon I_{A}]_+\otimes |k\rangle\langle k|.
$$
So, since the states $\rho$ and $\tilde{\sigma}$ commute and the (extended) quantum conditional entropy $S(A|B)$ defined by expression (\ref{ce-rep})
on the convex set $\S_{\mathrm{cq}}$ of all q-c states is a concave nonnegative function on
$\S_{\mathrm{cq}}$ satisfying inequality (\ref{LAA-2}) with $b_f=h_2$ and taking finite values at finite rank states in $\S_{\mathrm{cq}}$, Lemma \ref{b-lemma}B in Section 3 with $\S_0=\S_{\mathrm{cq}}$ implies that
$$
S(A|B)_{\sigma}=S(A|B)_{\hat{\sigma}}\geq S(A|B)_{\rho\ominus\varepsilon}-g(\varepsilon),
$$
while  Lemma \ref{b-lemma}A with $\S_0=\S_{\mathrm{cq}}$ implies that
$$
S(A|B)_{\sigma}=S(A|B)_{\hat{\sigma}}\geq S(A|B)_{\rho}-S(A|B)_{\rho\wedge\varepsilon}-g(\varepsilon)
$$
provided that $S(A|B)_{\rho}<+\infty$. $\Box$

\section{Characteristics of discrete random variables}

\subsection{On general methods of getting local lower bounds}

In this section we consider the class  of characteristics of discrete random variables $X_1$,...,$X_n$ each of which can be considered as a function $f$  on a convex subset
$\P_0$ of the set $\P_n$
of all $n$-variate probability distributions $\bar{p}=\{p_{i_1...i_n}\}_{(i_1,..,i_n)\in\N^n}$ (where $\N=\mathbb{N}\cup\{0\}$) taking values in $[0,+\infty]$ and satisfying
the inequalities
\begin{equation}\label{LAA-1-c}
  f(\lambda\bar{p}+(1-\lambda)\bar{q})\geq \lambda f(\bar{p})+(1-\lambda)f(\bar{q})-a_f(\lambda)
\end{equation}
and
\begin{equation}\label{LAA-2-c}
  f(\lambda\bar{p}+(1-\lambda)\bar{q})\leq \lambda f(\bar{p})+(1-\lambda)f(\bar{q})+b_f(\lambda)
\end{equation}
for any distributions  $\bar{p}$ and $\bar{q}$ in $\P_0$ and any $\lambda\in[0,1]$ with possible value $+\infty$ in both sides, where
$a_f(\lambda)$ and $b_f(\lambda)$ are continuous  functions on $[0,1]$  such that $a_f(0)=b_f(0)=0$. Inequalities  (\ref{LAA-1-c}) and (\ref{LAA-2-c})
are classical versions of inequalities (\ref{LAA-1}) and (\ref{LAA-2}). So, we may call functions satisfying (\ref{LAA-1-c}) and (\ref{LAA-2-c}) locally almost affine (briefly, LAA functions) on the set $\P_0$. As in Section 3 we will assume for technical simplicity that $a_f(\lambda)=a_f(1-\lambda)$ and $b_f(\lambda)=b_f(1-\lambda)$ for all $\lambda\in[0,1]$ and that
\begin{equation*}
 \textrm{ the functions}\;\; a_f\;\; \textrm{and}\;\; b_f\;\; \textrm{are concave and non-decreasing on}\;\; \textstyle[0,\frac{1}{2}].
\end{equation*}
The list of important entropic characteristics of discrete random variables satisfying inequalities (\ref{LAA-1-c}) and (\ref{LAA-2-c})
includes the Shannon (conditional) entropy, the mutual (conditional) information, the Kullback-Leibler divergence (as a function of the
first argument), the interaction information, etc. (see the subsections below and Section 4.4 in \cite{LCB}).

We will assume that  the set $\P_n$
of all $n$-variate probability distributions  is equipped with the total variance distance $\mathrm{TV}$ defined in (\ref{TVD}). Note that
many of the  characteristics mentioned before are lower semicontinuous functions
either on the whole set $\P_n$ or on proper subsets of $\P_n$ w.r.t. this distance. Below we will use the notation described in Section 2.2.

If a function $f$ defined on a subset $\P_0\subseteq\P_n$ is lower semicontinuous at a distribution  $\bar{p}\in \P_0$
then the quantity
\begin{equation*}
L_f(\bar{p}\shs|\shs\varepsilon)=\inf\left\{f(\bar{q})\,|\,\bar{q}\in\P_0,\, \textstyle\mathrm{TV}(\bar{p},\bar{q})\leq\varepsilon\right\}
\end{equation*}
tends to $f(\bar{p})\leq+\infty$ as $\,\varepsilon\to 0$.  We will say that
$\widehat{L}_f(\bar{p}\shs|\shs\varepsilon)$ is a \emph{faithful local lower bound} on a function $f$ at a distribution $\bar{p}$ if
$$
\widehat{L}_f(\bar{p}\shs|\shs\varepsilon)\leq L_f(\bar{p}\shs|\shs\varepsilon)\quad \forall\varepsilon\in(0,1]\quad\textrm{and}\quad \lim_{\varepsilon\to0}\widehat{L}_f(\bar{p}\shs|\shs\varepsilon)=f(\bar{p})\leq+\infty.
$$

Faithful local lower bound for  characteristics of discrete random variables satisfying inequalities (\ref{LAA-1-c}) and (\ref{LAA-2-c}) and such that
\begin{equation*}
|f(\bar{p})|\leq C_f\sum_{k=1}^m H(\bar{p}_{k})\leq+\infty,\quad C_f>0,\;\; m\leq n,
\end{equation*}
for any distribution  $\bar{p}$  in the domain of $f$ (where $\bar{p}_k$ denotes the marginal distribution of $\bar{p}$ corresponding to the $k$-th component and $H(\cdot)$ is the Shannon entropy)
can be obtained by using the one-side continuity bounds presented in Propositions 5 and 6 in \cite{LCB}. The concrete examples are considered in Sections 5.2 and 5.4 below.

Another way to get local lower bounds for characteristics of discrete random variables satisfying inequalities (\ref{LAA-1-c}) and (\ref{LAA-2-c}) is given by the
classical analogue of Lemma \ref{b-lemma} in Section 3 presented below.

For any $n$-variate probability distribution $\bar{p}\doteq\{p_{i_1...i_n}\}_{(i_1,..,i_n)\in\N^n}$ and arbitrary $\varepsilon$ in $(0,1]$ introduce two $n$-variate arrays of nonnegative numbers
\begin{equation}\label{2-op-c}
 \bar{p}\wedge\varepsilon \doteq \left\{\min\{p_{i_1...i_n},\varepsilon\}\right\}_{(i_1,..,i_n)\in\N^n},\quad
[\bar{p}-\varepsilon]_+\doteq \left\{\max\{p_{i_1...i_n}-\varepsilon,0\}\right\}_{(i_1,..,i_n)\in\N^n}.
\end{equation}
It is clear that $\,\bar{p}=\bar{p}\wedge\varepsilon+[\bar{p}-\varepsilon]_+$ and that $[\bar{p}-\varepsilon]_+$ tends to $\bar{p}$ as $\,\varepsilon\to0$.

Write $\widetilde{\P}_0$ for the cone generated by a subset $\P_0\subseteq \P_n$, i.e.
$\widetilde{\P}_0$  is the set of all $n$-variate arrays of nonnegative numbers of the form $\{cp_{i_1...i_n}\}$, $\{p_{i_1...i_n}\}\in\P_0$, $c\in\mathbb{R}_+$.

For any function $f$ on a convex subset $\P_0\subseteq\P_n$ denote by $\tilde{f}$ the  homogeneous extension of $f$ to the cone $\widetilde{\P}_0$  defined as
\begin{equation}\label{G-ext-c}
\tilde{f}(\bar{p})\doteq\|\bar{p}\|_1f(\bar{p}/\|\bar{p}\|_1), \quad \|\bar{p}\|_1\doteq\sum_{i_1,..,i_n}p_{i_1...i_n},
\end{equation}
for any nonzero array $\bar{p}=\{p_{i_1...i_n}\}$ in $\widetilde{\P}_0$ and equal to $0$ at the zero array.\smallskip

\begin{lemma}\label{b-lemma+} \emph{Let $f$ be a function on a convex subset $\,\P_0\subseteq\P_n$ taking values in $[0,+\infty]$ and satisfying inequalities (\ref{LAA-1-c}) and (\ref{LAA-2-c}) with possible values $+\infty$ in both sides. Let $\,\bar{p}\doteq\{p_{i_1...i_n}\}$ be a probability distribution in $\P_0$ with the properties
\begin{equation}\label{S-prop+c}
\!(\mathrm{a})\quad h(\bar{p})\in\widetilde{\P}_0\;\textrm{ for any  function }\;h:\mathbb{R}_+\to\mathbb{R}_+\textrm{ such that }\, h(x)\leq x\;\forall x\in\mathbb{R}_+,\qquad\qquad
\end{equation}
\begin{equation*}
\!(\mathrm{b})\quad  [\bar{p}-\bar{q}]_{\pm}=\{[p_{i_1...i_n}-q_{i_1...i_n}]_\pm\}\in\widetilde{\P}_0\;\textrm{ for any }\bar{q}\doteq\{q_{i_1...i_n}\}\textrm{ in }\,\P_0\,\textrm{ s.t. } \bar{q}\neq\bar{p},\qquad\quad\;
\end{equation*}
where  $h(\bar{p})=\{h(p_{i_1...i_n})\}$,  $\,[x]_-=\max\{-x,0\}$ and $\,[x]_+=\max\{x,0\}$.}\smallskip

\noindent A) \emph{If $f(\bar{p})<+\infty$ then
\begin{equation}\label{g-ob-c+0-c}
f(\bar{q})\geq f(\bar{p})-\tilde{f}(\bar{p}\wedge\varepsilon)-D_f(\varepsilon)-\tilde{a}_f(\varepsilon)
\end{equation}
for any probability distribution $\bar{q}$ in $\,\P_0$ such that $\,TV(\bar{p},\bar{q})\leq\varepsilon\leq1$,
where $\bar{p}\wedge\varepsilon$ is the array defined in (\ref{2-op-c}), $\tilde{f}$ is the homogeneous extension of $f$ defined in (\ref{G-ext-c}),}
\begin{equation*}
 D_f(\varepsilon)=\displaystyle(1+\varepsilon)(a_f+b_f)\!\left(\frac{\varepsilon}{1+\varepsilon}\right)\quad \textit{and}\quad\tilde{a}_f(\varepsilon)=\left\{\begin{array}{l}
        a_f(\varepsilon)\;\; \shs\textrm{if}\;\;  \varepsilon\in\shs[0,\frac{1}{2}]\\
        a_f(\frac{1}{2})\;\; \textrm{if}\;\;  \varepsilon\in(\frac{1}{2},1].
        \end{array}\right.
\end{equation*}\smallskip

\noindent B) \emph{If either $\,f(\bar{p})<+\infty\,$ or
\begin{equation}\label{f-a-c-c}
\tilde{f}([\bar{p}-\varepsilon]_+)<+\infty\quad \forall\varepsilon\in(0,1],
\end{equation}
where $[\bar{p}-\varepsilon]_+$ is the array defined in (\ref{2-op-c}),
then
\begin{equation}\label{g-ob-c+c}
f(\bar{q})\geq \tilde{f}([\bar{p}-\varepsilon]_+)-D_f(\varepsilon)-\tilde{a}_f(\varepsilon)-a_f(1-r_{\varepsilon})
\end{equation}
for any probability distribution $\bar{q}$ in $\,\P_0$ such that $\,TV(\bar{p},\bar{q})\leq\varepsilon\leq1$,
where $\,r_{\varepsilon}=\|[\bar{p}-\varepsilon]_+\|_1=\sum_{i_1,..,i_n}[p_{i_1...i_n}-\varepsilon]_+$.}
\end{lemma}\smallskip

\textbf{Remark 7.} All the terms in the r.h.s. of (\ref{g-ob-c+0-c}) and (\ref{g-ob-c+c}) are well defined and finite. Indeed,
the arrays $\bar{p}\wedge\varepsilon$ and $[\bar{p}-\varepsilon]_+$ belong to the cone $\widetilde{\P}_0$ due to condition (\ref{S-prop+c}).
If $\,f(\bar{p})<+\infty\,$ then inequality (\ref{LAA-1-c}) implies that $\tilde{f}(\bar{p}\wedge\varepsilon)<+\infty$ and $\tilde{f}([\bar{p}-\varepsilon]_+)<+\infty$, since  $\bar{p}\wedge\varepsilon+[\bar{p}-\varepsilon]_+=\bar{p}$.\smallskip

The lower bounds given by Lemma \ref{b-lemma+} can not be too accurate (by the reason described in Remark 2 in Section 3). The main advantage
of these lower bounds is their universality. Note also that the lower bound given by Lemma \ref{b-lemma+}B
is easily computable as the array  $[\bar{p}-\varepsilon]_+$ has a finite number of nonzero entries  for any $\bar{p}$ and $\varepsilon>0$.
\smallskip

\emph{Proof.} The set $\P_n$
of $n$-variate probability distributions $\{p_{i_1...i_n}\}_{(i_1,..,i_n)\in\N^n}$ is isomorphic to the set $\S_\tau$ of states in $\S(\H_{A_1...A_n})$ diagonizable in the basis
\begin{equation*}
 \{\tau^1_{i_1}\otimes\cdots\otimes\tau^n_{i_n}\}_{(i_1,..,i_n)\in\N^n},
\end{equation*}
where  $\{\tau^1_i\}_{i\in\N}$,...,$\{\tau^n_i\}_{i\in\N}\,$ are  orthonormal base in separable Hilbert spaces $\H_{A_1}$,.., $\H_{A_n}$
correspondingly. This isomorphism is given by the affine isometry
$$
\Theta:\bar{p}=\{p_{i_1...i_n}\}_{(i_1,..,i_n)\in\N^n}\;\mapsto\; \sum_{i_1,..,i_n}p_{i_1...i_n}|\tau^1_{i_1}\rangle\langle\tau^1_{i_1}|\otimes\cdots\otimes|\tau^n_{i_n}\rangle\langle\tau^n_{i_n}|
$$
from $\P_n$  onto $\S_\tau$. Since the set $\S_\tau$ consists of commuting states, the isomorphism $\Theta$ allows us to
derive all the claims of Lemma \ref{b-lemma+} from the corresponding claims of Lemma \ref{b-lemma} in Section 3. $\Box$

\subsection{The Shannon entropy and the equivocation}

The Shannon entropy $\,H(\bar{p})=-\sum_ip_i\ln p_i\,$ of $1$-variate probability distribution $\bar{p}=\{p_i\}\,$ is a concave lower semicontinuous function
on the set $\P_1$ taking values in $[0,+\infty]$ and satisfying inequality (\ref{w-k-ineq-c}). The lower semicontinuity of the Shannon entropy implies
that the quantity
\begin{equation*}
L_H(\bar{p}\shs|\shs\varepsilon)=\inf\left\{H(\bar{q})\,|\,\bar{q}\in\P_1,\, \mathrm{TV}(\bar{p},\bar{q})\leq\varepsilon\right\}
\end{equation*}
tends to $H(\bar{p})\leq+\infty\,$ as $\,\varepsilon\to+\infty$ for any probability distribution $\bar{p}$ in $\P_1$.

Since for any $\bar{p}=\{p_i\}$ and $\bar{q}=\{q_i\}$ we have
$$
H(\bar{p})=S(\rho), \quad H(\bar{q})=S(\sigma)\quad\textrm{ and }\quad \mathrm{TV}(\bar{p},\bar{q})=\textstyle\frac{1}{2}\|\rho-\sigma\|_1,
$$
where
$$
\rho=\sum_i p_i |\tau_i\rangle\langle \tau_i|\quad\textrm{ and }\quad \sigma=\sum_i q_i |\tau_i\rangle\langle \tau_i|
$$
are states on a separable Hilbert space $\H$ determined by a given orthonormal basic $\{\tau_i\}$ in $\H$, lower
bounds on  $L_H(\bar{p}\shs|\shs\varepsilon)$ can be obtained by "rewriting" the lower
bounds on  $L_S(\rho\shs|\shs\varepsilon)$ presented in Section 4.1. In particular, it follows from Proposition 1
that
\begin{equation}\label{B-lb-3+c}
 L_H(\bar{p}\shs|\shs\varepsilon)\geq H([\bar{p}-\varepsilon]_+)-g(\varepsilon)
\end{equation}
for any $\bar{p}$ in $\P_1$, where $g$ is the function defined in (\ref{g-def}) and
\begin{equation*}\label{sums+c}
H([\bar{p}-\varepsilon]_+)\doteq c_{\varepsilon}H(c^{-1}_{\varepsilon}[\bar{p}-\varepsilon]_+)=\sum_{i\in I_\varepsilon} \eta(p_i-\varepsilon)-\eta(c_{\varepsilon}),\quad c_{\varepsilon}=\|[\bar{p}-\varepsilon]_+\|_1=\sum_{i\in I_\varepsilon}(p_i-\varepsilon),
\end{equation*}
$I_\varepsilon$ is the \emph{finite} set of all $i$ such that $p_i-\varepsilon>0$. The lower bound (\ref{B-lb-3+c}) is faithful: its r.h.s. tends to $\,H(\bar{p})\leq+\infty\,$ as $\,\varepsilon\to0$.\smallskip

\textbf{Example 4.} Consider the probability distribution
$\bar{p}_*=\left\{(ci\ln^2i)^{-1}\right\}_{i=5}^{+\infty}$, where $c=\sum_{i=5}^{+\infty} \frac{1}{i\ln^2i}<1$,
with infinite Shannon entropy. Lower bound (\ref{B-lb-3+c}) and the estimates made in Example 2 in Section 4.1 show that
$$
L_H(\bar{p}_*\shs|\shs\varepsilon)\geq\frac{1}{c}\ln  \frac{-\ln(c\varepsilon)}{3\ln5}-\frac{1}{ce}+o(1)=O(\ln(-\ln\varepsilon))\quad \textrm{as}\;\; \varepsilon\to0.
$$
This is a lower bound on the \emph{unknown} rate of convergence of $L_H(\bar{p}_*\shs|\shs\varepsilon)$ to $H(\bar{p}_*)=+\infty$ as $\varepsilon\to0$.

\medskip

The \emph{Shannon conditional entropy} $H(X_1|X_2)$ (also called \emph{equivocation}) of random variables $X_1$ and $X_2$ with joint probability distribution $\bar{p}=\{p_{i_1i_2}\}$ is defined as
$$
H(X_1|X_2)_{\bar{p}}= \sum_{i_1,i_2} p_{i_1i_2}\!\left(-\ln\frac{p_{i_1i_2}}{[\bar{p}_2]_{i_2}}\right)=H(\bar{p})-H(\bar{p}_2),
$$
where the second formula is valid if $H(\bar{p}_2)$ is finite (while the first one is well defined for any $\bar{p}$ in $\P_2$, since all the terms in the series are nonnegative) \cite{C&T}. The function $\bar{p}\mapsto H(X_1|X_2)_{\bar{p}}$ is lower semicontinuous on the set $\P_2$ and takes values in $[0,+\infty]$. It satisfies inequalities  (\ref{LAA-1-c}) and (\ref{LAA-2-c}) with $a_f=0$ and $b_f=h_2$.\footnote{Inequality (\ref{LAA-1-c}) with $a_f=0$ means concavity of the conditional entropy,
inequality (\ref{LAA-2-c}) with $b_f=h_2$ can be proved easily for probability distributions $\bar{p}$ and $\bar{q}$ such that $H(\bar{p}_2)$ and $H(\bar{q}_2)$ are finite
by using concavity of the Shannon entropy and inequality (\ref{w-k-ineq-c}). The validity of this inequality in the general case can be proved by approximation.}
The lower semicontinuity of this function  implies that the quantity
\begin{equation*}
L_{H(X_1|X_2)}(\bar{p}\shs|\shs\varepsilon)=\inf\left\{H(X_1|X_2)_{\bar{q}}\,|\,\bar{q}\in\P_2,\, \mathrm{TV}(\bar{p},\bar{q})\leq\varepsilon\right\}
\end{equation*}
tends to $H(X_1|X_2)_{\bar{p}}\leq+\infty\,$ as $\,\varepsilon\to+\infty$ for any $\bar{p}\in\P_2$. Below we will obtain several lower bounds on $L_{H(X_1|X_2)}(\bar{p}\shs|\shs\varepsilon)$.

The well known inequality
$$
H(X_1|X_2)_{\bar{p}}\leq H(\bar{p}_1)
$$
and the validity of inequalities (\ref{LAA-1-c}) and (\ref{LAA-2-c}) for the function $\bar{p}\mapsto H(X_1|X_2)_{\bar{p}}$ with $a_f=0$ and $b_f=h_2$ mentioned before
imply that this function belongs to the class $T_2^1(1,1)$ in the notation used in Section 4.4 in \cite{LCB}.

Thus, if $\,\bar{p}\,$ is such that the marginal distribution $\bar{p}_1$ has a finite number of nonzero entries (denoted by $|\bar{p}_1|$)  then Proposition 5B in \cite{LCB} implies that
\begin{equation}\label{CE-CB-1-c}
H(X_1|X_2)_{\bar{p}}-H(X_1|X_2)_{\bar{q}}\leq \varepsilon\ln|\bar{p}_1|+g(\varepsilon)
\end{equation}
for any $\bar{q}$ in $\P_2$ such that $\,\mathrm{TV}(\bar{p},\bar{q})\leq\varepsilon$, where $g$ is the function defined in (\ref{g-def}). This shows that
\begin{equation}\label{CE-LB-c-1}
L_{H(X_1|X_2)}(\bar{p}\shs|\shs\varepsilon)\geq H(X_1|X_2)_{\bar{p}}- \varepsilon\ln|\bar{p}_1|-g(\varepsilon)\quad \forall\varepsilon\in(0,1].
\end{equation}

If $\,\bar{p}\,$ is such that $\sum_{i=0}^{+\infty}E_i[\bar{p}_1]_i=E<+\infty$, where $\{E_i\}_{i=0}^{+\infty}$ is a nondecreasing sequence of nonnegative numbers with $E_0=0$ satisfying condition (\ref{Z-cond}), then  Proposition 6B in \cite{LCB} implies that
\begin{equation}\label{CE-CB-2-c}
H(X_1|X_2)_{\bar{p}}-H(X_1|X_2)_{\bar{q}}\leq \varepsilon F_{\SC}(E/\varepsilon)+g(\varepsilon)
\end{equation}
for any $\bar{q}$ in $\P_2$ such that $\,\mathrm{TV}(\bar{p},\bar{q})\leq\varepsilon$, where $F_{\SC}$ is the function defined in (\ref{F-Z}) with $\SC=\{E_i\}_{i=0}^{+\infty}$.  This shows that
\begin{equation}\label{CE-LB-c-2}
L_{H(X_1|X_2)}(\bar{p}\shs|\shs\varepsilon)\geq H(X_1|X_2)_{\bar{p}}-\varepsilon F_{\SC}(E/\varepsilon)-g(\varepsilon)\quad \forall\varepsilon\in(0,1].
\end{equation}
The equivalence of (\ref{Z-cond}) and (\ref{Z-cond+}) imply that the r.h.s. of (\ref{CE-LB-c-2}) tends to $H(X_1|X_2)_{\bar{p}}$ as $\varepsilon\to0$.

One-side continuity bounds (\ref{CE-CB-1-c}) and (\ref{CE-CB-2-c}) are analysed in Example 2 in \cite[Section 4.4]{LCB}, where it is shown that
they are asymptotically tight for large $|\bar{p}_1|$ and $E$ correspondingly. Thus, lower bounds (\ref{CE-LB-c-1}) and (\ref{CE-LB-c-2}) are faithful and quite accurate (provided that $\bar{p}$ satisfies the corresponding conditions). By Proposition 4
in \cite{EC} the existence of a nondecreasing sequence $\{E_i\}_{i=0}^{+\infty}$ of nonnegative numbers satisfying condition (\ref{Z-cond}) such that
$\sum_{i=0}^{+\infty} E_i[\bar{p}_1]_i<+\infty$ is equivalent to the finiteness of the Shannon entropy of $\bar{p}_1=\{[\bar{p}_1]_i\}_i$. So,
lower bound (\ref{CE-LB-c-2}) is applicable if $H(\bar{p}_1)$ is finite.\smallskip

Not too accurate but universal and simple lower bounds on $L_{H(X_1|X_2)}(\bar{p}\shs|\shs\varepsilon)$ can be obtained by applying Lemma \ref{b-lemma+} to the function
$f(\bar{p})=H(X_1|X_2)_{\bar{p}}$.\smallskip

\textbf{Proposition 6.} \emph{Let $\,\widetilde{H}(X_1|X_2)$ be
the homogeneous extension of $H(X_1|X_2)$ to the cone of $\,2$-variate arrays of nonnegative numbers defined according to the rule (\ref{G-ext-c}).}\smallskip

\noindent A) \emph{If $\,\bar{p}$ is an arbitrary probability distribution in $\P_2$ then
\begin{equation}\label{CE-LB++c}
L_{H(X_1|X_2)}(\bar{p}\shs|\shs\varepsilon)\geq \widetilde{H}(X_1|X_2)_{[\bar{p}-\varepsilon]_+}-g(\varepsilon)\quad \textstyle\forall\varepsilon\in(0,1],
\end{equation}
where $[\bar{p}-\varepsilon]_+$ is the array defined in (\ref{2-op-c}).}\smallskip

\noindent B) \emph{If $\,\bar{p}$ is a probability distribution in $\P_2$ such that $H(X_1|X_2)_{\bar{p}}<+\infty$ then
\begin{equation}\label{CE-LB+c}
L_{H(X_1|X_2)}(\bar{p}\shs|\shs\varepsilon)\geq H(X_1|X_2)_{\bar{p}}-\widetilde{H}(X_1|X_2)_{\bar{p}\wedge\varepsilon}-g(\varepsilon)\quad \textstyle\forall\varepsilon\in(0,1],
\end{equation}
where $\bar{p}\wedge\varepsilon$ is the array defined in (\ref{2-op-c}).}\smallskip

\emph{The lower bounds (\ref{CE-LB++c}) and (\ref{CE-LB+c}) are faithful: their r.h.s.tend to $\,H(X_1|X_2)_{\bar{p}}\leq+\infty\,$ as $\,\varepsilon\to+\infty$.}
\smallskip

\emph{Proof.} The main claims of the proposition follow directly from Lemma \ref{b-lemma+} with $\P_0=\P_2$ and the properties of the
function $\bar{p}\mapsto H(X_1|X_2)_{\bar{p}}$ mentioned before. We have only to note
that condition (\ref{f-a-c-c}) holds for this function, since it takes finite values at any probability distributions in $\P_2$
with a finite number of nonzero entries.\smallskip

To prove the last claim it suffices to note that $\widetilde{H}(X_1|X_2)_{[\bar{p}-\varepsilon]_+}$ tends to $H(X_1|X_2)_{\bar{p}}$ as $\varepsilon\to0$ due to the lower semicontinuity of the function $\bar{q}\mapsto \widetilde{H}(X_1|X_2)_{\bar{q}}$  and the inequality
\begin{equation}\label{CE-ineq-c}
H(X_1|X_2)_{\bar{p}}\geq \widetilde{H}(X_1|X_2)_{\bar{p}\wedge\varepsilon}+\widetilde{H}(X_1|X_2)_{[\bar{p}-\varepsilon]_+}
\end{equation}
which follows from the concavity of this function, since $\,\bar{p}=\bar{p}\wedge\varepsilon+[\bar{p}-\varepsilon]_+$. $\Box$
\smallskip

\textbf{Remark 8.}
Inequality (\ref{CE-ineq-c}) implies that lower bound (\ref{CE-LB+c}) is sharper than  (\ref{CE-LB++c}). The advantage of lower bound (\ref{CE-LB++c})
consists in its validity for \emph{all} probability distributions $\bar{p}$ (including probability distributions with infinite $H(X_1|X_2)_{\bar{p}}$). Note also
that the r.h.s. of (\ref{CE-LB++c}) is easily calculated, since the array $[\bar{p}-\varepsilon]_+$ contains a finite number of nonzero entries.

\subsection{Affine functionals and the Kullback-Leibler divergence}

Let $\mathcal{S}=\{E_i\}_{i=0}^{+\infty}$ be a sequence of nonnegative numbers. Then
$$
E_{\mathcal{S}}(\bar{p})=\sum_{i=0}^{+\infty} E_ip_i,\quad \bar{p}=\{p_i\}_{i=0}^{+\infty},
$$
is an affine lower semicontinuous function  on the set $\P_1$ taking
values in $[0,+\infty)$. The lower semicontinuity of $E_{\mathcal{S}}$ implies that for any $\bar{p}\in\P_1$ the quantity
\begin{equation}\label{E-fun+}
L_{E_{\mathcal{S}}}(\bar{p}\shs|\shs\varepsilon)=\inf\left\{E_{\mathcal{S}}(\bar{q})\,|\,\bar{q}\in\P_1,\, \mathrm{TV}(\bar{p},\bar{q})\leq\varepsilon\right\}
\end{equation}
tends to $E_{\mathcal{S}}(\bar{p})\leq+\infty\,$ as $\,\varepsilon\to0^+$. Direct application of Lemma \ref{b-lemma+} gives the following\smallskip

\textbf{Proposition 7.}  \emph{Let $\mathcal{S}=\{E_i\}_{i=0}^{+\infty}$ be a sequence of nonnegative numbers and $\bar{p}=\{p_i\}_{i=0}^{+\infty}$ a probability distribution in $\P_1$. Then
\begin{equation}\label{H-LB+c}
 L_{E_{\mathcal{S}}}(\bar{p}\shs|\shs\varepsilon)\geq \sum_{i\in I_\varepsilon}E_i(p_i-\varepsilon)\quad \forall\varepsilon\in(0,1],
\end{equation}
where $I_\varepsilon$ is the set of all $i$ such that $p_i>\varepsilon$.}

\emph{The lower bound (\ref{H-LB+c}) is faithful: its  r.h.s. tends to $\,E_{\mathcal{S}}(\bar{p})\leq+\infty\,$ as $\,\varepsilon\to0$.}\smallskip

\textbf{Note:} Since the set $I_\varepsilon$ is finite, the r.h.s. of (\ref{H-LB+c}) is easily calculated for any $\varepsilon>0$.\smallskip

\textbf{Example 5.} Assume that $\mathcal{S}=\{E_i\}_{i=0}^{+\infty}$ is a nondecreasing sequence with $E_0=0$  and $\bar{p}=\{p_i\}_{i=0}^{+\infty}$ is a degenerate distribution with $p_j=1$ for some $j>0$.
Then it is easy to see that the infimum in (\ref{E-fun+}) is attained at the distribution  $\,\bar{q}=\varepsilon\{1,0,0,...\}+(1-\varepsilon)\bar{p}\,$ for any $\varepsilon\in(0,1]$ and hence both sides of (\ref{H-LB+c}) are equal to $E_j(1-\varepsilon)$.\medskip

The \emph{Kullback-Leibler divergence} for probability distributions $\bar{p}=\{p_i\}$ and $\bar{q}=\{q_i\}$ is defined as
\begin{equation*}
  D_{KL}(\bar{p}\shs\|\bar{q})=\sum_i p_i\ln (p_i/q_i),
\end{equation*}
where it is
assumed that $0\ln0=0$ and $D_{KL}(\bar{p}\shs\|\bar{q})=+\infty$ if $\,\supp \bar{p}\,$ is not contained in $\,\supp \bar{q}\,$ \cite{K&L,C&T}.\footnote{The support $\,\supp \bar{p}\,$ of a probability distribution $\bar{p}=\{p_i\}$ is the set of all $i$ such that $p_i>0$.}  For any given $\bar{q}$ the function $\,D_{\bar{q}}(\bar{p})\doteq D_{KL}(\bar{p}\shs\|\bar{q})\,$
is lower semicontinuous on $\P_1$,  takes values in $[0,+\infty]$ and satisfies inequalities  (\ref{LAA-1-c}) and (\ref{LAA-2-c}) with $a_f=h_2$ and $b_f=0$ \cite{K&L,C&T,O&P}.
The lower semicontinuity of this function implies
that for any $\bar{p}$ and $\bar{q}$ the quantity
\begin{equation*}
L_{D_{\bar{q}}}(\bar{p}\shs|\shs\varepsilon)=\inf\left\{D_{KL}(\bar{r}\shs\|\bar{q})\,|\,\bar{r}\in\P_1,\, \mathrm{TV}(\bar{p},\bar{r})\leq\varepsilon\right\}
\end{equation*}
tends to $\,D_{KL}(\bar{p}\shs\|\bar{q})\leq+\infty\,$ as $\,\varepsilon\to+\infty$.

For any natural $d$ consider the quantity
\begin{equation*}
L^{d}_{D_{\bar{q}}}(\bar{p}\shs|\shs\varepsilon)=\inf\left\{D_{KL}(\bar{r}\shs\|\bar{q})\,|\,\bar{r}\in\P_1,\, |\bar{r}|\leq d, \mathrm{TV}(\bar{p},\bar{r})\leq\varepsilon\right\},
\end{equation*}
where it is assumed that $\inf \emptyset=+\infty$. This quantity is an upper bound on $L_{D_{\bar{q}}}(\bar{p}\shs|\shs\varepsilon)$. It also tends to $\,D_{KL}(\bar{p}\shs\|\bar{q})\leq+\infty\,$ as $\,\varepsilon\to+\infty$ provided that $|\bar{p}|\leq d$. If $|\bar{q}|<+\infty$ then it is easy to see that  $L_{D_{\bar{q}}}(\bar{p}\shs|\shs\varepsilon)=L^{d}_{D_{\bar{q}}}(\bar{p}\shs|\shs\varepsilon)$ for all $d\geq|\bar{q}|$.\smallskip


B using Lemma \ref{b-lemma+} and Proposition 7 one can prove the following\smallskip

\textbf{Proposition 8.} \emph{Let $\bar{p}=\{p_i\}$ and $\bar{q}=\{q_i\}$ be probability distributions in $\P_1$ such that $\,\supp \bar{p}\subseteq\supp\bar{q}$.} \emph{Then
\begin{equation}\label{KLD-LB+}
L_{D_{\bar{q}}}(\bar{p}\shs|\shs\varepsilon)\geq \sum_{i\in I_\varepsilon}(p_i-\varepsilon)\ln\frac{p_i-\varepsilon}{c_{\varepsilon}q_i} -g(\varepsilon)-\tilde{h}_2(\varepsilon)-h_2(1-c_\varepsilon)\quad  \forall\varepsilon\in(0,1],
\end{equation}
where $I_\varepsilon$ is the set of all $\,i\shs$ such that $\,p_i>\varepsilon$, $\,c_{\varepsilon}=\sum_{i\in I_\varepsilon}(p_{i}-\varepsilon)\,$ and
$\,\tilde{h}_2$ is the modification of the binary entropy  defined in (\ref{h+}).}\smallskip

\emph{If $\,H(\bar{p})<+\infty$ and $\,d\geq2\,$ is a natural number then}
\begin{equation}\label{KLD-LB+d}
\!L^{d}_{D_{\bar{q}}}(\bar{p}\shs|\shs\varepsilon)\geq \sum_{i\in I_\varepsilon}(-\ln q_i)(p_i-\varepsilon)-H(\bar{p})-\varepsilon\ln(d-1)-h_2(\varepsilon)\quad\forall\varepsilon\in(0,1\!-\!1/d].\!
\end{equation}

\emph{The lower bounds (\ref{KLD-LB+}) and (\ref{KLD-LB+d}) are faithful: their r.h.s. tend to $\,D_{KL}(\bar{p}\shs\|\bar{q})\leq+\infty\,$ as $\,\varepsilon\to+\infty$.}
\smallskip

\textbf{Remark 9.}
Since the set $I_\varepsilon$ is finite for any $\varepsilon>0$, the right hand sides of (\ref{KLD-LB+}) and  (\ref{KLD-LB+d}) are finite and easily calculated.

The condition $\,\supp \bar{p}\subseteq\supp\bar{q}\,$ in Proposition 8 is essential. If $\,\supp \bar{p}\nsubseteq\supp\bar{q}\,$
then it is easy to see that $\,L_{D_{\bar{q}}}(\bar{p}\shs|\shs\varepsilon)=L^d_{D_{\bar{q}}}(\bar{p}\shs|\shs\varepsilon)=+\infty\,$ for all sufficiently small $\varepsilon>0$.

\smallskip

\emph{Proof.}  The function $\,D_{\bar{q}}(\bar{p})=D_{KL}(\bar{p}\shs\|\shs\bar{q})\,$ is nonnegative convex and satisfies inequality (\ref{LAA-1-c}) with $a_f=h_2$. It is easy to see that condition (\ref{f-a-c-c}) with $f=D_{\bar{q}}$ is valid for any probability distribution $\bar{p}$ such that $\,\supp \bar{p}\subseteq\supp\bar{q}$. Thus, Lemma \ref{b-lemma+}B with $\P_0=\P_1$ shows that
$$
D_{\bar{q}}(\bar{r})\geq c_{\varepsilon}D_{\bar{q}}(c^{-1}_{\varepsilon}[\bar{p}-\varepsilon]_+)-g(\varepsilon)-\tilde{h}_2(\varepsilon)-h_2(1-c_\varepsilon),\quad  c_{\varepsilon}\doteq\|[\bar{p}-\varepsilon]_+\|_1,
$$
for any probability distribution $\bar{r}$ in $\P_1$ such that $\mathrm{TV}(\bar{p},\bar{r})\leq\varepsilon\leq1$, where $[\bar{p}-\varepsilon]_+$ is the array defined in (\ref{2-op-c}). This
implies the lower bound (\ref{KLD-LB+}).

 Assume that $\bar{r}$ is  a probability distribution in $\P_1$ such that $|\bar{r}|\leq d$ and $\mathrm{TV}(\bar{p},\bar{r})\leq\varepsilon\leq1-1/d$.
Then $H(\bar{r})<+\infty$ and hence
$$
D(\bar{r}\|\shs\bar{q})=\sum_{i=0}^{+\infty}(-\ln q_i)r_i-H(\bar{r})=\sum_{i=0}^{+\infty}(-\ln q_i)r_i-H(\bar{p})+[H(\bar{p})-H(\bar{r})].
$$
The classical version of Proposition 2 in \cite{LCB} shows that
\begin{equation*}
H(\bar{p})-H(\bar{r})\geq -\varepsilon\ln(d-1)-h_2(\varepsilon)
\end{equation*}
(regardless of the value of $|\bar{p}|$). Thus, Proposition 7 implies that
$$
D(\bar{r}\|\shs\bar{q})\geq\sum_{i\in I_{\varepsilon}}(-\ln q_i)(p_i-\varepsilon)-H(\bar{p})-\varepsilon\ln(d-1)-h_2(\varepsilon).
$$

To prove the last claim of the proposition it suffices to note that
$$
\lim_{\varepsilon\to0}c_{\varepsilon}D(c^{-1}_{\varepsilon}[\bar{p}-\varepsilon]_+\|\shs\bar{q})=D(\bar{p}\shs\|\shs\bar{q})\leq+\infty
$$
and that
$$
\lim_{\varepsilon\to0}\sum_{i\in I_{\varepsilon}}(-\ln q_i)(p_i-\varepsilon)=\sum_{i=0}^{+\infty}(-\ln q_i)p_i\leq+\infty.
$$
The second limit relation is obvious, the first one can be easily shown by using the lower semicontinuity of the nonnegative function $\,D_{\bar{q}}(\bar{p})=D_{KL}(\bar{p}\shs\|\shs\bar{q})\,$
and  the validity of  inequality (\ref{LAA-1-c}) for this function with $a_f=h_2$ mentioned before. $\Box$

\subsection{The mutual information}

The \emph{mutual information} of random variables $X_1$ and $X_2$ with joint probability distribution $\bar{p}=\{p_{i_1i_2}\}$ is defined as
\begin{equation*}\label{MI-def}
I(X_1\!:\!X_2)_{\bar{p}}=D_{KL}(\bar{p}\shs\|\bar{p}_1\times\bar{p}_2)=H(\bar{p}_1)+H(\bar{p}_2)-H(\bar{p}),
\end{equation*}
where the second formula is valid  provided that $H(\bar{p})$ is finite. By the properties of the Kullback-Leibler divergence the function $\bar{p}\mapsto I(X_1\!:\!X_2)_{\bar{p}}$ is lower semicontinuous on $\P_2$
and takes values in $[0,+\infty]$. It satisfies inequalities (\ref{LAA-1-c}) and (\ref{LAA-2-c}) with $a_f=b_f=h_2$ with possible value $+\infty$ in both sides.\footnote{This follows, in particular, from the inequalities (86) and (87) in \cite{QC}.}
The lower semicontinuity of this function implies
that for any $\bar{p}\in\P_2$ the quantity
\begin{equation*}
L_{I(X_1\!:X_2)}(\bar{p}\shs|\shs\varepsilon)=\inf\left\{I(X_1\!:\!X_2)_{\bar{q}}\,|\,\bar{q}\in\P_2,\, \mathrm{TV}(\bar{p},\bar{q})\leq\varepsilon\right\}
\end{equation*}
tends to $\,I(X_1\!:\!X_2)_{\bar{p}}\leq+\infty\,$ as $\,\varepsilon\to+\infty$. Below we will obtain several faithful lower bounds on $L_{I(X_1\!:X_2)}(\bar{p}\shs|\shs\varepsilon)$.

The well known inequality
$$
I(X_1\!:\!X_2)_{\bar{p}}\leq H(\bar{p}_1)
$$
and the validity of inequalities (\ref{LAA-1-c}) and (\ref{LAA-2-c}) for the function $\bar{p}\mapsto I(X_1\!:\!X_2)_{\bar{p}}$ with $a_f=b_f=h_2$ mentioned before
imply that this function belongs to the class $T_2^1(1,2)$ in the notation used in Section 4.4 in \cite{LCB}.

Thus, if $\bar{p}$ is such that either $|\bar{p}_1|$ or $|\bar{p}_2|$ is finite then Proposition 5B in \cite{LCB} and the symmetry arguments imply that
\begin{equation}\label{I-CB-1}
I(X_1\!:\!X_2)_{\bar{p}}-I(X_1\!:\!X_2)_{\bar{q}}\leq \varepsilon\ln\min\{|\bar{p}_1|,|\bar{p}_2|\}+2g(\varepsilon)
\end{equation}
for any $\bar{q}$ in $\P_2$ such that $\,\mathrm{TV}(\bar{p},\bar{q})\leq\varepsilon$  ($|\bar{p}_k|$ denotes the number of nonzero entries of the marginal distribution $\bar{p}_k$, $g(\varepsilon)$ is the function defined in (\ref{g-def})).  This shows that
\begin{equation}\label{I-LB-c-1}
L_{I(X_1\!:X_2)}(\bar{p}\shs|\shs\varepsilon)\geq I(X_1\!:\!X_2)_{\bar{p}}- \varepsilon\ln\min\{|\bar{p}_1|,|\bar{p}_2|\}-2g(\varepsilon)\quad \forall\varepsilon\in(0,1].
\end{equation}

If $\bar{p}$ is such that $\sum_{i=0}^{+\infty} E_i[\bar{p}_k]_i=E<+\infty$, where $k$ is either $1$ or $2$ and $\{E_i\}_{i=0}^{+\infty}$ is a nondecreasing sequence of nonnegative numbers with $E_0=0$ satisfying condition (\ref{Z-cond}), then  Proposition 6B in \cite{LCB} and the symmetry arguments imply that
\begin{equation}\label{I-CB-2}
I(X_1\!:\!X_2)_{\bar{p}}-I(X_1\!:\!X_2)_{\bar{q}}\leq \varepsilon F_{\SC}(E/\varepsilon)+2g(\varepsilon)
\end{equation}
for any $\bar{q}$ in $\P_2$ such that $\,\mathrm{TV}(\bar{p},\bar{q})\leq\varepsilon$, where $F_{\SC}$ is the function defined in (\ref{F-Z}) with $\SC=\{E_i\}_{i=0}^{+\infty}$.  This shows that
\begin{equation}\label{I-LB-c-2}
L_{I(X_1\!:X_2)}(\bar{p}\shs|\shs\varepsilon)\geq I(X_1\!:\!X_2)_{\bar{p}}-\varepsilon F_{\SC}(E/\varepsilon)-2g(\varepsilon)\quad \forall\varepsilon\in(0,1].
\end{equation}
The equivalence of (\ref{Z-cond}) and (\ref{Z-cond+}) imply that the r.h.s. of (\ref{I-LB-c-2}) tends to $I(X_1\!:\!X_2)_{\bar{p}}\,$ as $\,\varepsilon\to0$.

Thus, both lower bounds (\ref{I-LB-c-1}) and (\ref{I-LB-c-2}) are faithful (provided that $\bar{p}$ satisfies the corresponding conditions). By Proposition 4
in \cite{EC} the existence of a nondecreasing sequence $\{E_i\}_{i=0}^{+\infty}$ of nonnegative numbers satisfying condition (\ref{Z-cond}) such that\break
$\,\sum_{i=0}^{+\infty} E_i[\bar{p}_k]_i<+\infty\,$ is equivalent to the finiteness of the Shannon entropy of the distribution $\{[\bar{p}_k]_i\}_i$. So,
the lower bound (\ref{I-LB-c-2}) is applicable if either $H(\bar{p}_1)$ or $H(\bar{p}_2)$ is finite.

Coarse but universal and easily computable lower bounds on $L_{I(X_1\!:X_2)}(\bar{p}\shs|\shs\varepsilon)$ can be obtained by applying Lemma \ref{b-lemma+} to the function
$f(\bar{p})=I(X_1\!:\!X_2)_{\bar{p}}$.\smallskip

\textbf{Proposition 9.} \emph{Let $\widetilde{I}(X_1\!:\!X_2)$ be
the homogeneous extension of $I(X_1\!:\!X_2)$ to the cone of $\,2$-variate arrays of nonnegative numbers defined according to the rule (\ref{G-ext-c}).}

\noindent A) \emph{If $\,\bar{p}$ is an arbitrary probability distribution in $\P_2$ then
\begin{equation}\label{I-LB++}
L_{I(X_1\!:X_2)}(\bar{p}\shs|\shs\varepsilon)\geq \widetilde{I}(X_1\!:\!X_2)_{[\bar{p}-\varepsilon]_+}-2g(\varepsilon)-\tilde{h}_2(\varepsilon)-h_2(1-r_\varepsilon)\quad \textstyle\forall\varepsilon\in(0,1],
\end{equation}
where $[\bar{p}-\varepsilon]_+$ is the array defined in (\ref{2-op-c}), $\,r_\varepsilon=\|[\bar{p}-\varepsilon]_+\|_1=\sum_{i_1,i_2}[p_{i_1i_2}-\varepsilon]_+$ and $\tilde{h}_2$ is the function defined in (\ref{h+}).}\smallskip

\noindent B) \emph{If $\,\bar{p}$ is a probability distribution in $\P_2$ such that $I(X_1\!:\!X_2)_{\bar{p}}<+\infty$ then
\begin{equation}\label{I-LB+}
L_{I(X_1\!:X_2)}(\bar{p}\shs|\shs\varepsilon)\geq I(X_1\!:\!X_2)_{\bar{p}}-\widetilde{I}(X_1\!:\!X_2)_{\bar{p}\wedge\varepsilon}-2g(\varepsilon)-\tilde{h}_2(\varepsilon)\quad \textstyle\forall\varepsilon\in(0,1],
\end{equation}
where $\bar{p}\wedge\varepsilon$ is the array defined in (\ref{2-op-c}).}\smallskip

\emph{The lower bounds (\ref{I-LB++}) and (\ref{I-LB+}) are faithful: their r.h.s. tend to $\,I(X_1\!:\!X_2)_{\bar{p}}\leq+\infty\,$ as $\,\varepsilon\to+\infty$.}
\smallskip

\emph{Proof.} The main claims of the proposition follow directly from Lemma \ref{b-lemma+} with $\P_0=\P_2$ and the properties of the
function $\bar{p}\mapsto I(X_1\!:\!X_2)_{\bar{p}}$ mentioned at the begin of this subsection. We have only to note
that condition (\ref{f-a-c-c}) holds for this function, since it takes finite values at any probability distributions in $\P_2$
with a finite number of nonzero entries.

To prove the last claim it suffices to note that $\widetilde{I}(X_1\!:\!X_2)_{[\bar{p}-\varepsilon]_+}$ tends to $I(X_1\!:\!X_2)_{\bar{p}}$ as $\varepsilon\to0$
due to the lower semicontinuity of the function $\bar{p}\mapsto \widetilde{I}(X_1\!:\!X_2)_{\bar{p}}$ and the inequality
\begin{equation}\label{I-ineq}
I(X_1\!:\!X_2)_{\bar{p}}\geq \widetilde{I}(X_1\!:\!X_2)_{\bar{p}\wedge\varepsilon}+ \widetilde{I}(X_1\!:\!X_2)_{[\bar{p}-\varepsilon]_+}-h_2(r_{\varepsilon}),
\end{equation}
which follows from the validity of inequality (\ref{LAA-1-c}) for this function with $a_f=h_2$ mentioned before,
since  $\bar{p}=\bar{p}\wedge\varepsilon+[\bar{p}-\varepsilon]_+$. $\Box$\smallskip

\textbf{Remark 10.}
Inequality (\ref{I-ineq}) implies that lower bound (\ref{I-LB+}) is sharper than  (\ref{I-LB++}). The advantage of lower bound (\ref{I-LB++})
consists in its validity for all probability distributions $\bar{p}$ (including probability distributions $\bar{p}$ with infinite $I(X_1\!:\!X_2)_{\bar{p}}$). Note also
that the r.h.s. of (\ref{I-LB++}) is easily calculated, since the array $[\bar{p}-\varepsilon]_+$ contains a finite number of nonzero entries.

\bigskip

I am grateful to A.S.Holevo and to the participants of his seminar
"Quantum probability, statistic, information" (the Steklov
Mathematical Institute) for useful discussion.

\end{document}